\documentclass[a4paper,11pt]{article}
\pdfoutput=1 

\usepackage{jcappub} 
\usepackage[utf8]{inputenc}
\usepackage[T1]{fontenc} 
\usepackage{graphicx}
\usepackage{epsfig}
\usepackage{rotate}
\usepackage{amsmath}
\usepackage{amssymb}
\usepackage{amsfonts}
\usepackage{bm}
\usepackage{enumerate}
\usepackage{xcolor} 
\usepackage{natbib, hyperref}
\usepackage{upgreek}
\usepackage{empheq}

\newcommand{\ud}{\mathrm{d}}

\newcommand{\qoo}{\mathbb{Q}_{o}}
\newcommand{\hs}{\mathbb{H}}
\newcommand{\ho}{\hs}
\newcommand{\xo}{\mathbb{X}}

\newcommand{\hsa}{\langle\mathbb{H}\rangle}

\newcommand{\thsa}{\langle\tilde{\mathbb{H}}\rangle}

\newcommand{\ih}{\mathbb{I}}
\newcommand{\tiha}{\langle\tilde{\mathbb{I}}\rangle}

\newcommand{\xoa}{\langle\mathbb{X}\rangle}

\newcommand{\tho}{\tilde{\mathbb{H}}}

\newcommand{\hoo}{\mathbb{H}_{\o}}

\newcommand{\xoo}{\mathbb{X}_{\o}}

\newcommand{\thoo}{\tilde{\mathbb{H}}_{\o}}

\newcommand{\txo}{\tilde{\mathbb{X}}}

\newcommand{\txoo}{\tilde{\mathbb{X}}_{\o}}

\newcommand{\tqoo}{\tilde{\mathbb{Q}}_{\o}}

\def\tvn{\tilde{{n}}}

\def\tn{\tilde{n}}
\def\tv{\tilde{v}}
\def\tu{\tilde{u}}

\def\tK{\tilde{K}}

\def\tz{\tilde{z}}

\def\d{{\rm d}}

\def\m{\rm m}

\def\o{\rm o}
\def\e{\rm s}

\def\be{\begin{equation}}
\def\ee{\end{equation}}
\def\bea{\begin{eqnarray}}
\def\eea{\end{eqnarray}}

\title{Covariant cosmography:\\
the observer-dependence of the Hubble parameter}
\vspace*{-2.5cm}
\author{Roy Maartens$^{1,2,3}$, Jessica Santiago$^{4,5}$, 
Chris Clarkson$^{6,1,7}$,\\ Basheer Kalbouneh$^8$, Christian Marinoni$^8$}
 
\affiliation{\footnotesize{$^1$Department of Physics \& Astronomy, University of the Western Cape, Cape Town 7535, South Africa\\
$^2$Institute of Cosmology \& Gravitation, University of Portsmouth, Portsmouth PO1 3FX, United Kingdom \\
$^{3}$National Institute for Theoretical \& Computational Sciences, Cape Town 7535, South Africa\\
$^4$Leung Center for Cosmology \& Particle Astrophysics,
National Taiwan University, Taipei 10617, Taiwan
\\$^5$Section of Astrophysics, Astronomy \& Mechanics, Department of Physics,
Aristotle University of Thessaloniki, Thessaloniki 54124, Greece
\\$^{6}$School of Physics \& Astronomy, Queen Mary University of London, London E1 4NS, United Kingdom \\
 $^7$Department of Mathematics \& Applied Mathematics, University of Cape Town, Cape Town 7701, South Africa\\
$^8$Aix Marseille Univ, Universit\'e de Toulon, CNRS, CPT, Marseille, France
 }}

\abstract{
The disagreement between low- and high-redshift measurements of the Hubble parameter is emerging as a serious challenge to the standard model of cosmology.  We develop a covariant cosmographic analysis  of the Hubble parameter in a general spacetime, which is fully model-independent and can thus be used as part of a robust assessment of the tension. Here our focus is not on the tension but on  understanding the relation between the physical expansion rate and its measurement by observers -- which is critical for model-independent measurements and tests.
We define the physical Hubble parameter and its multipoles in a general spacetime and derive for the first time the covariant boost transformation of the multipoles measured by a heliocentric observer.  The analysis is extended to the covariant deceleration parameter.
Current cosmographic measurements of the expansion anisotropy contain discrepancies
and disagreements, some of which may arise because the correct transformations
for a moving observer are not applied. 
A heliocentric observer will detect a dipole, generated not only by a Doppler effect, but also by an aberration effect due to shear. In principle, the observer can measure both the intrinsic shear anisotropy and the velocity of the observer relative to the matter -- without any knowledge of peculiar velocities, 
which are gauge dependent and do not arise in a covariant approach.
The practical implementation of these results is investigated in a follow-up paper.
We further show that the standard cosmographic relation between the Hubble parameter, the redshift and the luminosity distance (or magnitude) is {\em not} invariant under boosts and holds only in the matter frame. A moving observer who applies the standard cosmographic relation should correct the luminosity distance by a redshift factor  -- otherwise an incorrect dipole and a spurious octupole are predicted. }

\begin{document}
\maketitle
\flushbottom

\section{Introduction}

After more than 100 years, the Friedman-Lemaître-Robertson-Walker (FLRW) spacetime remains the basis for the standard cosmological model. 
The fundamental parameters of this model have been measured with good precision by recent surveys -- and the next generation of cosmological surveys promises spectacular improvements in precision. However, advances in measurements are also exposing previously hidden puzzles and problems.  Some anomalies or tensions  are emerging as a challenge to the standard model (see e.g. \cite{ Perivolaropoulos:2021jda, Abdalla:2022yfr,Aluri:2022hzs,Peebles:2022akh,Verde:2023lmm} for recent reviews). This is exacerbated by the still-unknown nature of dark energy and dark matter.
 
Challenges to the standard model are to be expected, and even welcomed, in cosmology: they are not only problems, but also opportunities for new advances. Probably the most serious cosmic tension arises from the disagreement between the low- and high-redshift values of the Hubble parameter $H_0$. There are many observational and theoretical aspects to this tension, which has spurred a new wave of activity that is attempting to explain and resolve it \cite{ Perivolaropoulos:2021jda,Abdalla:2022yfr,Aluri:2022hzs,Peebles:2022akh,Verde:2023lmm}. 

In this paper, our focus is not on the Hubble tension itself, but rather on  the need for a model-independent framework to analyse the Hubble parameter and its measurement at low redshifts in a general expanding universe. Why is this necessary? Precisely because the Hubble tension and related tensions challenge the foundations of the standard model. A model-independent approach that does not assume an FLRW or any other spacetime metric, allows us to  see the challenge from a more general viewpoint and to draw more robust conclusions.

In \autoref{sec2}, we give a covariant analysis of  the matter frame and lightrays  in a general spacetime, and we derive the covariant transformations of redshifts and directions under a change of observer. We present in
\autoref{sec3} the covariant definition of the Hubble parameter and its  multipolar decomposition. Then we derive for the first time the transformation of multipoles measured by a moving observer, whose motion induces a dipole from a combined Doppler and aberration effect. The result indicates how in principle the moving observer can measure simultaneously the intrinsic anisotropy of the expansion rate and the observer's velocity relative to the matter frame.  We highlight the fact that peculiar velocities of galaxies, which are a gauge-dependent feature of the standard cosmological model, do not arise in a covariant cosmography. This allows us to extract the anisotropy and the observer's velocity without any peculiar velocity information. Practical application of this result is taken up in the follow-up paper \cite{Kalbouneh:2024tfw}.
We also make a comparison with the standard approach in a perturbed FLRW model and show how the covariant  Hubble parameter, when specialised to perturbed FLRW, produces a gauge-invariant expression.
In \autoref{sec4} 
we derive the covariant cosmographic relation between cosmic distances and the Hubble parameter in the matter frame. 
We show that the redshift-distance relation is invariant under a boost only for the area distance. By contrast, the redshift-luminosity distance (or magnitude) relation is not boost-invariant. Consequently, a moving observer should use a corrected luminosity distance -- otherwise an incorrect dipole and a spurious octupole arise.
In  FLRW and perturbed FLRW models, the  error leads to a false prediction of {\em no} dipole.
Then we briefly discuss the potential bias on measurement from local structure and the contamination from the covariant deceleration parameter (and higher-order jerk and curvature parameters). This topic is investigated in detail in the follow-up paper \cite{Kalbouneh:2024tfw}. We summarise our results and conclude in \autoref{sec5}.

\section{Matter, observers and lightrays}
\label{sec2}

In a covariant approach to observational cosmology, the spacetime metric and the coordinates (or tetrad vectors) remain general and no perturbative approximation is needed  (for reviews, see e.g. \cite{Ellis:1998ct,Tsagas:2007yx,Clarkson:2010uz,Ellis:2012}). A physical choice is made of a 4-velocity vector field $u^a$, and then a 1+3 splitting is performed, using the projection tensor into the instantaneous rest-space of $u^a$  at each event:
\begin{align}\label{hdef}
   \boxed{ h^{ab}=g^{ab}+u^a u^b \quad \mbox{with}\quad u^au_a=-1\,,~~h_{ab}u^b=0\,.}
\end{align}

\subsection{Matter frame in a general spacetime}

In our case, the obvious
physical choice of 4-velocity is that of the matter, $u_{\rm m}^a$, whose flowlines encode the expansion rate.
On cosmological scales at late times, 
it is standard to model the matter as a pressure-free fluid, i.e. as `dust', which is invariantly defined by its energy-momentum tensor
$T_{\rm m}^{ab}$. The dust 4-velocity is then a smooth vector field defined uniquely by the vanishing of momentum flux (see \cite{Ellis:2012}, chap. 5):
\begin{align}\label{m4vel}
\boxed{q_{\rm m}^a=-u_{\rm m}^c\,h^a_{{\rm m}\,b}\, 
T_{{\rm m}\,c}^b =0 \quad \Rightarrow\quad T_{\rm m}^{ab} =\rho_{\rm m}\,u_{\rm m}^a\, u_{\rm m}^b \,,} 
\end{align}
where $\rho_{\rm m}$ is the physical matter density.
This defines a physical cosmic frame 
--  i.e. the frame that  moves with the matter and thus shares the matter  4-velocity  $u_{\rm m}^a$.
In the special case of perturbed FLRW, the matter frame corresponds to the comoving gauge. However, there is {\em no} gauge choice, {\em no} background spacetime, and {\em no} perturbative expansion involved in the covariant matter frame.
Note that the matter energy-momentum tensor is an invariant physical tensor defined only by the dust matter, independent of observers. The matter-frame observer measures the physical matter density $\rho_{\m}$ and vanishing pressure and momentum flux. A moving observer $\tu^a$ will measure a different density, $\tilde\rho=T^{ab}_{\m}\tu_a \tu_b$, together with nonzero pressures, $\tilde P$ and $\tilde\pi_{ab}$, and momentum flux, $\tilde q_a$, which arise purely from relative velocity -- see Appendix~\ref{appa} for further details.

Vanishing pressure  leads via  momentum conservation \eqref{momcg}
to vanishing 4-acceleration:
\bea \label{acc}
\dot{u}_{\rm m}^a= u_{\rm m}^b\nabla_b u_{\rm m}^a =0.
\eea
It is also reasonable to neglect vorticity on cosmological scales: 
\begin{align}\label{vort}
\omega_{\m}^{ab}= h^{[a}_{{\m}\, c} h^{b]}_{{\m}\, d} \nabla^c u_{\rm m}^d=0\,,    
\end{align}
where square brackets denote the anti-symmetric part. 
The expansion tensor of matter flowlines is then
\bea\label{ukin}
 \boxed{\nabla^a u_{\rm m}^b  
 = \frac{1}{3}\Theta_{\rm m}\, h_{\rm m}^{ab}+\sigma_{\rm m}^{ab}\equiv \Theta_{\rm m}^{ab} \quad \mbox{where}~~\Theta_{\rm m}={\nabla}_a u_{\rm m}^a\,,~~\sigma_{\rm m}^{ab}= h_{{\rm m}\,c}^{\langle a} h_{{\rm m}\,d}^{b\rangle} \nabla^c u_{\rm m}^d\,.}
\eea
Here $\Theta_{\rm m}$ is the volume expansion rate of flowlines, and  the shear rate is 
$\sigma_{\rm m}^{ab}=\sigma_{\rm m}^{\langle ab \rangle}$, 
where 
angled brackets indicate the projected,  symmetric, tracefree part. This is defined in general, relative to some 4-velocity $u^a$ with projection tensor $h_{ab}$, by 
\bea \label{stfs}
W_{\langle ab \rangle} = \Big[h_{a}^{(c} h_{b}^{d)}-\frac{1}{3} h_{ab}h^{cd} \Big]W_{cd}\,.
\eea
We note that\\

\fbox{\parbox{0.9\textwidth}{\em The matter expansion  tensor is an invariant physical tensor defined only by the dust matter, independent of observers.}}\\

\noindent Matter-frame observers measure the physical expansion and shear rates. Moving observers with 4-velocity $\tu^a$ will measure different expansion and shear rates, which involve the relative velocity -- as we show in \autoref{secbhp}. 

Another physical frame in a general spacetime is the cosmic microwave background (CMB) frame,  defined covariantly by the 4-velocity $u_{\rm cmb}^a$ relative to which the dipole of the photon distribution function $\mathsf{f}$ vanishes \cite{1983AnPhy.150..455E,ETM1983,Ellis:2012,Stoeger:1994qs,Tsagas:2007yx, Clarkson:2010uz}: 
\bea \label{cmbd}
F^a \equiv\frac{3}{4\pi}\int \ud \Omega_{e}\, \mathsf{f}(x^b,p^c)\, e^a =0 \quad \mbox{where}\quad e_a e^a=1\,,~~e_a u_{\rm cmb}^a=0\,.
\eea
Here $e^a$ is the direction of the  photon 4-momentum $p^a=E(u_{\rm cmb}^a+e^a)$, and $\ud \Omega_{e}$ is the solid angle element about $e^a$. 
The condition $F^a=0$ for a vanishing dipole implies a vanishing radiation flux \cite{1983AnPhy.150..455E,ETM1983}: $q^a_{\rm cmb}\equiv (4\pi/3)\int \ud E\,E^3F^a=0$, similar to the matter-frame case \eqref{m4vel}.

In practice the CMB frame is straightforward to determine, given the precision of CMB data and the absence of nonlinear complications. By contrast, the matter frame is very difficult to measure. However, as we show in \autoref{secbhp}, we do not need to know the matter frame a priori in order to determine the Hubble rate -- indeed, the heliocentric observer's velocity relative to the matter frame can be simultaneously determined, as we show (see also \cite{Kalbouneh:2024tfw}).

FLRW models obey the Cosmological Principle exactly: they are exactly isotropic and homogeneous in the unique cosmic frame defined by the 4-velocity field $\bar{u}^a$ that is normal to the 3-surfaces of homogeneity and isotropy. As a result, all physical 4-velocity fields must coincide: $\bar u_{\rm cmb}^a=\bar u_{\rm m}^a= \bar{u}^a$. 
Perturbed FLRW models obey the Cosmological Principle in the more general sense, i.e. they are statistically homogeneous and isotropic, so that the correlations of perturbations on the background 3-surfaces are invariant under translation and rotation. 
This implies that the matter and CMB 4-velocities must agree up to small perturbations, i.e.
\begin{align}\label{cosprin}
u_{\rm cmb}^a=\bar{u}^a+ \delta u_{\rm cmb}^a\,,\quad 
u_{\rm m}^a=\bar{u}^a+ \delta u_{\rm m}^a\,.
\end{align}
Testing for statistically significant differences between $u_{\rm m}^a$ and $u_{\rm cmb}^a$ is a powerful test of the Cosmological Principle, as argued by \cite{ellis1984}. 

Here we do {\em not} assume the Cosmological Principle, since our aim is to investigate properties of the Hubble parameter in a general spacetime.

\subsection{Observers and lightrays}

Consider a matter-frame observer O with 4-velocity $u^a_{\m}|_{\o}$ and a boosted  observer $\tilde{\rm O}$ with 4-velocity $\tu^a|_{\o}$, both at the event o of cosmic observations (`here and now' in cosmic space and time). For our purposes the
boosted observer  at o is typically a heliocentric observer, whose  4-velocity differs from the matter 4-velocity at o. The reason for this difference is as follows. The averaging scale which defines the dust 4-velocity is larger than a galaxy -- galaxies are treated as particles, in free fall under gravity. The
heliocentric  observer is also in free fall -- but within a galaxy, so that its motion is governed by local dynamics on scales that are well below the dust averaging scale.  

The boosted observer $\tilde{\rm O}$ at o  moves with instantaneous velocity $v_{\o}^a$ relative to O:
\be \label{6}
\boxed{\tu^a\circeq \gamma\big(u_{\rm m}^a+v^a\big) \circeq  u_{\rm m}^a+v^a+ O(v^2)
\quad\mbox{with}\quad \gamma \circeq \big(1-v^2\big)^{-1/2}\,,~~ v_a\,u_{\rm m}^a  \circeq 0 \,,}
\ee
where we introduced the shorthand notation
\bea
X \circeq Y ~~\Leftrightarrow~~ X\big|_{\o}=Y\big|_{\o} \,.
\eea
The inverse relation of \eqref{6} is \cite{Maartens:1998xg}
\bea \label{6x}
 u_{\rm m}^a \circeq \gamma\big(\tu^a+\tilde{v} ^a\big) 
\quad\mbox{where}\quad
\tilde{v}^a \circeq - \gamma\big(v^a+v^2 u_{\rm m}^a \big) 
~~ \mbox{with} ~~\tu_a \tilde{v}^a \circeq 0, ~~\tilde{v}^2\circeq v^2\,.
\eea
Here $\tv^a_{\o}$ is the velocity of O measured by $\tilde{\rm O}$.
The boosted  projection tensor follows from \eqref{6} as 
\bea \label{tildh}
\boxed{\tilde{h}^{ab}\circeq h_{\rm m}^{ab}+\gamma^2\Big(v^2 u_{\rm m}^au_{\rm m}^b+u_{\rm m}^{a}v^{b}+u_{\rm m}^{b}v^{a}+v^a v^b \Big).}
\eea
As expected, the projection of $v^a$ into the boosted rest-space is proportional to $-\tilde v^a$:
\begin{align} \label{tvv}
 \tilde{h}^a_b\, v^b \circeq - \gamma\, \tilde v^a\,.  
\end{align}
To leading order in $v_{\o}$, we have
\bea
u_{\rm m}^a\circeq  \tu^a-{v}^a+ O({v}^2)\,,\quad  \tilde{v}^a \circeq -v^a+ O({v}^2),
\quad 
\tilde{h}^{ab}\circeq h_{\rm m}^{ab}+u_{\rm m}^{a}v^{b}+u_{\rm m}^{b}v^{a}+ O({v}^2).~~
\eea

The above relations hold at an arbitrary event, but
the boosted observer is only needed at the cosmic here-and-now event o: {\em no information beyond {\rm o} is involved in the boost transformations at} o -- and no  congruence of boosted observers is implied or required.
This covariant scenario changes in perturbed FLRW models, where the Cosmological Principle implies a congruence of boosted observers whose worldlines cross the 3-surface $t=t_0$ which contains the event o. (See \autoref{sec2.2}.)

\begin{figure}[!ht]
\centering
\includegraphics[width=0.45\textwidth]{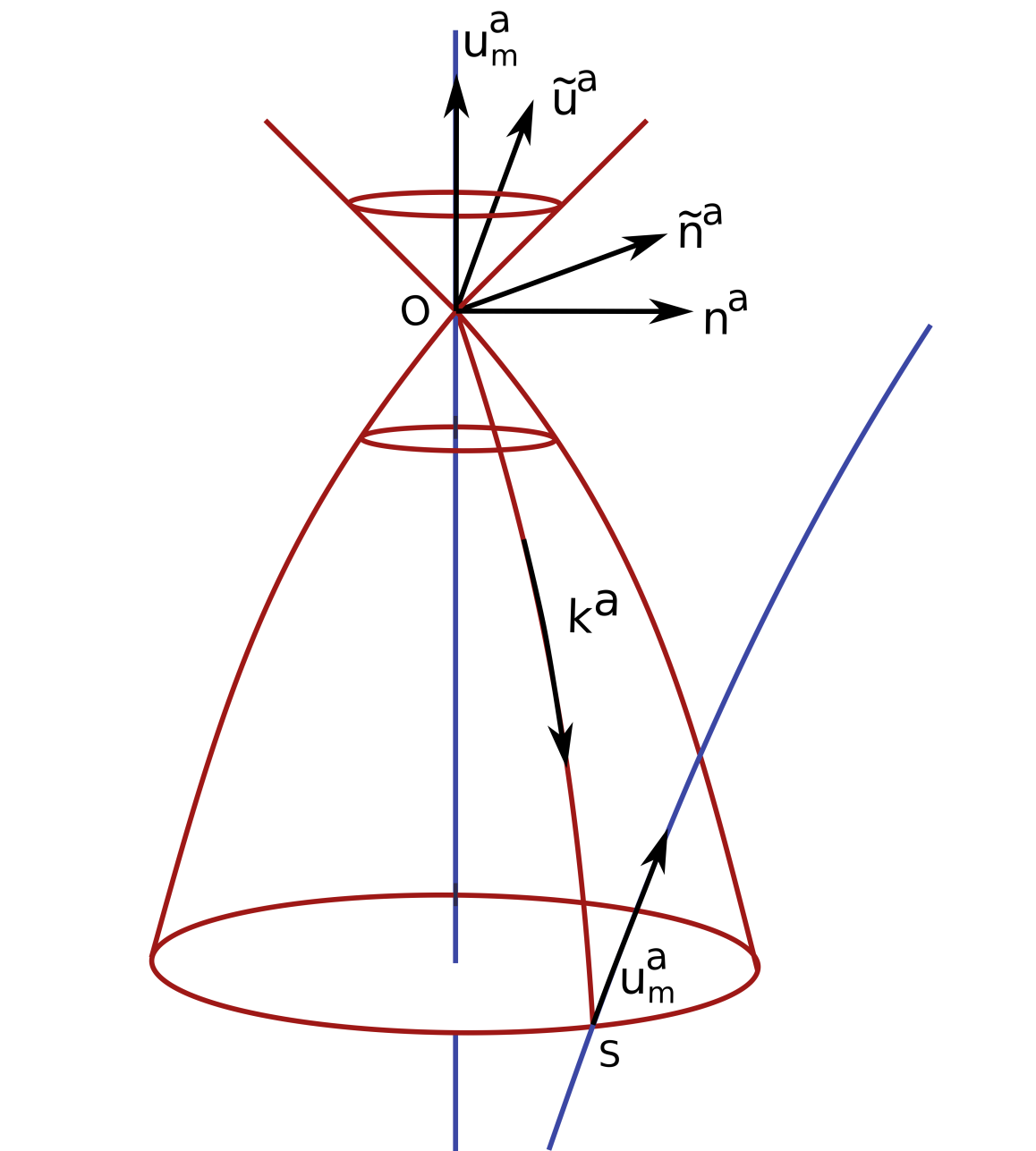} \\
\caption{The key 4-vectors: 4-velocities of the matter  observer O and moving observer $\tilde{\rm O}$ at event o, and of the matter source at event s;  direction 4-vectors of the lightray relative to O and $\tilde{\rm O}$ at o; lightray 4-vector.}
 \label{fig0}
\end{figure}

At the event o,
any observer receives information along lightrays of the past lightcone of the event o. Lightrays are described by the photon 4-momentum $p^a$. For an observer-centred approach, it is useful to use the reverse (past-pointing) 4-vector $k^a$. In the matter frame, we can decompose it as
\be \label{12}
\boxed{k^a \equiv -p^a
  = E\big( -u_{\rm m}^a+n^a\big)
 \,,\quad E=-p_au_{\rm m}^a=k_a u_{\rm m}^a\,,~ u_{\rm m}^an_a=0\,,~n^a n_a=1\,,}
\ee
where $E$ is the photon energy measured by  O and  $n^a$ is the  line-of-sight direction of the lightray as seen by O (see \autoref{fig0}). \\

\fbox{\parbox{0.9\textwidth}{
$k^a$ {\em is an invariant physical 4-vector field defined purely by the past lightcone, independent of observers. But its energy and direction   are observer-dependent:} 
\vspace*{-.2cm}
\bea \label{keun}
k^a\circeq E( -u_{\rm m}^a+n^a)\circeq \tilde{k}^a \circeq \tilde E( -\tu^a+\tn^a)\,.
\eea
\vspace*{-.5cm}}}\\

\noindent The redshift of a matter source at event s, with matter 4-velocity $u_{\rm m}^a|_{\e}$, depends on the  observer's 4-velocity at o. The redshift measured in the matter frame at o is given by 
\be \label{15}
\boxed{1+z\equiv \frac{(k_a u_{\rm m}^a)_{\e}} {(k_a u_{\rm m}^a)_{\o}}= \frac{E_{\e}}{E_{\o}}\,.}
\ee\\
{\em This redshift  is a physical quantity that is defined  by the dust matter and the past lightcone.} A moving observer will measure a different redshift, according to a covariant boost transformation (see \autoref{sec2.3c}).  Note that there is no `peculiar velocity' of the source in the  covariant cosmography; this point is further discussed in \autoref{sec2.2}.
 
\subsection{Covariant boost transformations}
\label{sec2.3c}

By \eqref{keun}, the lightray 4-momentum is observer-independent, $\tilde k^a\circeq k^a$.  
We use this relation to find the covariant Doppler and aberration effects as follows.\\

\noindent$\bullet$ We project \eqref{keun} along $\tu^a|_{\o}$: 
\bea
k^a\tu_a \circeq \tilde E \circeq E(-u_{\rm m}^a+n^a)\gamma(u^{\rm m}_a+v_a)\,,
\eea
which leads to the Doppler shifted energy measured by  $\tilde{\rm O}$: 
\bea
\tilde E &\circeq&  \Gamma E\circeq \gamma \big(1+v_a n^a \big) E \quad \mbox{where}\quad
\Gamma \equiv  \gamma \big(1+v_a n^a \big)
\\ &\circeq & \big(1+v_a n^a \big) E+ O(v^2)
\,. \notag
\eea
If we make explicit the dependence on directions, then
\bea
\boxed{\tilde E(\tvn) \circeq  \Gamma(n) E(n)
\quad \mbox{where}\quad\Gamma(n) =  \gamma \big(1+v_a n^a \big) 
\,.} \label{elb}
\eea
Note that the inverse transformation to \eqref{elb} is
\bea
 \boxed{E(n) \circeq  \tilde\Gamma(\tvn) \tilde E(\tvn)\quad \mbox{where}\quad
\tilde\Gamma(\tvn) =  \gamma \big(1+\tilde{v}_a \tvn^a \big) = \Gamma(n)^{-1}
\,.} \label{elbi}
\eea
$\bullet$ We project \eqref{keun} into the rest-space of $\tilde{\rm O}$ using \eqref{tildh}, to find that 
\bea
k^b\,\tilde{h}^a_b\circeq\tilde E \, \tn^a
\circeq E\Big[n^a+ \gamma^2\big(v^2u_{\rm m}^a+v^a \big) 
+\gamma^2\,v_b n^b\big(u_{\rm m}^a+v^a \big)\Big].
\eea
Using \eqref{elb}, this leads to the aberrated direction  measured by  $\tilde{\rm O}$:
\begin{align}
    \label{nlb}
\Aboxed{\tn^a &\circeq \Gamma^{-1}
\Big[n^a+ \gamma\Gamma\,
v^a +\gamma^2
\big(v^2+v_b n^b \big)u_{\rm m}^a \Big]} 
\\ \notag
&\circeq \big(1- v_bn^b \big)n^a+v^a+v_bn^b\,u_{\rm m}^a+ O(v^2)\,,
\end{align} 
which agrees with the result in \cite{Maartens:1998xg}. Note that $\tn^a$ in \eqref{nlb} necessarily has a nonzero component along $u_{\rm m}^a$ -- since $\tn^a$ is in the rest-space of $\tilde{\rm O}$, not O, as illustrated in \autoref{fig0}. 

The above covariant relations at event o reduce to the standard Lorentz boost relations in inertial coordinates, in accordance with  the Equivalence Principle. The coordinate-independent derivation of the covariant boost transformations is arguably simpler and more clear than the derivation in special relativity with two inertial coordinate systems. 

In order to remove dependence of the lightray vector on the photon energy $E_{\o}$ at event o, we can define a normalised lightray vector at o:
\bea \label{nork}
 \boxed{K^a \circeq \frac{k^a}{E} \circeq -u_{\rm m}^a+n^a\,.}
\eea
Under a boost it transforms as:
\begin{align} \label{tkkb}
\Aboxed{\tK^a & \circeq  \Gamma(n)^{-1} K^a \circeq \gamma^{-1}\big(1+v_bn^b \big)^{-1} K^a }\\ \notag
&\circeq  \big(1-v_bn^b \big) K^a + O(v^2)\,,
\end{align} 
by \eqref{6} and \eqref{nlb}. Unlike the photon 4-momentum, $K^a$ is not invariant.    
Finally, the redshifts measured by O and $\tilde{\rm O}$ of a matter source at event s, are related by
\be\label{15x}
\frac{1+\tilde z}{1+z} = \frac{E_{\e}}{\tilde E_{\o}} \frac{E_{\o}}{ E_{\e}}=\frac{E_{\o}}{\tilde E_{\o}}\,,
\ee
which follows from \eqref{15}, using the fact that  the photon energy in the matter frame  at the source, $E_{\e}$, is independent of the observers at o. 
Together with \eqref{elb}, this gives
\begin{align}
    \label{15y}
\Aboxed{1+\tilde z(\tvn) &= \Gamma_{\o}(n)^{-1} \big[1+z(n)\big]}
\\ \notag
&= \big(1-v_an^a \big)_{\o}\big[1+z(n)\big]+ O(v^2_{\o}) \,.
\end{align}

\section{Covariant Hubble parameter and its multipoles}
\label{sec3}

The basis for a covariant definition of the Hubble parameter is the matter expansion tensor \eqref{ukin}, which contains information about the behaviour of neighbouring dust matter flowlines in a general
spacetime. 
Since $\Theta_{\rm m}^{ab}\,u_{{\rm m}\,b}=0$, this information is in the matter rest-space. In order to construct an (in principle) {\em observable} Hubble parameter, we need to project the information onto the observer's past lightcone.

Following the pioneering papers \cite{Kristian:1966}, \cite{Ellis:1971pg}\footnote{Based on 1969 lectures; reprinted in \cite{ellis2009}.} and \cite{MacCallum:1970}, we define the covariant Hubble parameter $\hs_{\o}$ in a general spacetime as:\footnote{See also \cite{Maartens:1980,Ellis:1985,Clarkson:1999zq,Clarkson:2010uz,Umeh:2013, Heinesen:2020bej,Umeh:2022prn}.}
\bea  \label{14}
\boxed{\hs   
\circeq \Theta_{\rm m}^{ab}\, K_a K_b
\circeq K_a K_b  \,\nabla^a u_{\rm m}^b 
\quad\mbox{where} \quad 
K^a \circeq  -u_{\rm m}^a+n^a \,.}
\eea
Here we project not along $k^a$, but its normalised form $K^a$, defined in \eqref{nork}, in order to remove dependence on the photon energy at the observer, $E_{\o}$. 
We highlight the point that
:\\ 

\fbox{\parbox{0.9\textwidth}{
$\hs_{\o}$  {\em is a physical quantity that is defined  by the dust matter and the past lightcone.}}} \\

\noindent 
In practice, the Hubble parameter is typically measured via distance estimates.
Although $\hs_{\o}$ has been defined independently of distances, we show in \autoref{sec4} that
\begin{align}
\boxed{\hoo = \frac{\ud z}{\ud d} \bigg|_{\o},}
\end{align}
where $d$ is any covariant cosmic distance, measured in the matter frame.

The physical $\hs_{\o}$ is measured by a matter-frame observer. A moving observer will measure a different $\thoo$, according to the covariant boost transformation derived  in \autoref{secbhp} below.
From \eqref{ukin} and \eqref{14} we find the directional dependence of the Hubble parameter in the multipolar form (see Appendix~\ref{appcovm} for some details on covariant multipoles):
\be \label{14xx}
\boxed{\hs (n) \circeq \hsa+ \hs_{ab}\, n^{\langle a} n^{b\rangle}
\quad \mbox{where}~~ \hsa=\frac{1}{3}\Theta_{\rm m} \,,~~ \hs^{ab}=\sigma_{\rm m}^{ab}=\hs^{\langle ab \rangle}
 \,.}
\ee
Equation \eqref{14xx} shows that in any spacetime with dust matter, there is no dipole in the physical Hubble parameter -- i.e. measured in the  matter frame -- only a monopole and quadrupole.
{\em   A dipole in the Hubble parameter means  that the observer is moving relative to the matter frame.}
This provides another covariant way, alternative to \eqref{m4vel}, to define the matter frame in principle as {\em the frame in which the Hubble parameter has no dipole}.

 For the simplest case, i.e. the spatially homogeneous and isotropic standard cosmology, the quadrupole vanishes and \eqref{14xx} reduces to:
\begin{align}
\hs(n)_{\o}= {\hsa}_{\o}= H_0  \quad \mbox{in FLRW}, 
\end{align} 
where o is any event in the 3-surface $t=t_0$.

\subsection{Boosted Hubble parameter}
\label{secbhp}

The matter expansion tensor is a physical tensor independent of observers and is therefore unaffected by a change of observer at event o. 
However the  Hubble parameter defined by \eqref{14} does depend on the observer, since $\tilde K^a\neq K^a$. Only a matter-frame observer measures the physical Hubble parameter. 
For a heliocentric or other moving observer, the covariant boost transformation of the Hubble parameter  follows from the transformation of $K^a$:
\begin{align} \label{14t}
\tho(\tvn)&\circeq {\Theta}_{\rm m}^{ab}\, \tilde K_a \tilde K_b 
\circeq {\Theta}_{\rm m}^{ab}\, 
  \Big(\gamma^2 v_av_b-2\gamma v_b\tilde n_a+\tilde n_a\tilde n_b\Big) \notag\\
&\circeq {\Theta}_{\rm m}^{ab}\,\Big[ \Big(\gamma^2 v_av_b+\frac{1}{3}\tilde{h}_{ab}\Big)-
\big(2\gamma v_b \big)\tilde n_a+\tilde n_{\langle a}\tilde n_{b\rangle}\Big],
\end{align}
where we used \eqref{6}. The second line of \eqref{14t} makes clear the multipolar structure of $\thoo$, which can also be checked by the direct method
of \eqref{a7} and \eqref{a8}. 
The multipole expansion seen by the moving observer, with direction vector $\tn^a$, is 
\begin{align}
\tho(\tvn) \circeq 
\thsa +\tilde{\hs}_a\,\tn^a+ \tilde{\hs}_{\langle ab\rangle}\,\tn^{ a} \tn^{b}\,,
\label{bhpobs}
\end{align}
where
\begin{subequations}
\begin{empheq}[box=\fbox]{align}
\thsa &\circeq 
\Big(1 +{\frac{4}{3}\gamma^2v^2\Big)\hsa + \frac{4}{3}\gamma^2 \sigma_{\rm m}^{ab}\,{v}_a {v}_b} 
 \circeq \hsa +O(v^2),  
\label{thhm} \\
\tilde \hs^a &\circeq -2\gamma \Big( \hsa \,v^b+\sigma_{\rm m}^{bc}\,{v}_c \Big) \tilde{h}^a_{b}
\qquad\quad \, \circeq -2 \Big( \hsa \,v^a+\sigma_{\rm m}^{ab}\,{v}_b \Big) +O(v^2) ,
\label{thhd}
\\
\tilde \hs^{ab} &\circeq  \tilde{h}^{\langle a}_c \tilde{h}^{b\rangle}_d \,{\Theta}_{\rm m}^{cd}
\qquad\qquad\qquad\qquad\quad \, \circeq \sigma_{\rm m}^{ab}+O(v^2)
\,.
\label{thhq}
\end{empheq}
\end{subequations}
The explicit covariant form of the  Hubble parameter measured by a boosted observer
is a new result, to our knowledge. In \cite{Heinesen:2020bej}, the luminosity distance $d_L$ is expanded in a redshift series whose coefficients $d_L^{(i)}$ depend on $\ho^{-1}_{\o}$ (and higher-order cosmographic parameters). Then  the boosted coefficients $\tilde d_L^{(i)}$ are presented in terms of $\tho^{-1}_{\o}$,   but without investigating the effect of the boost on the Hubble multipoles. The explicit forms of the boosted Hubble parameter and higher-order parameters turn out be crucial for covariant cosmography, as we show in this section and in \autoref{sec3.2} (see also \cite{Kalbouneh:2024tfw}).

It is essential that the heliocentric multipoles are defined relative to $\tn^a$, which is the line of sight that is used in observations -- unlike the matter-frame direction $n^a$ which is not known by a heliocentric observer.
Note also  that the projections and tracefree parts of
vectors and tensors above
are defined in terms of the heliocentric observer's projection tensor $\tilde h^{ab}$, not the matter-frame projector $h_{\m}^{ab}$. 
In summary, at leading order in velocity: \\

\fbox{\parbox{0.92\textwidth}{
\begin{itemize} \vspace*{-0.2cm}
\item
{\em The boosted monopole and quadrupole  equal the physical Hubble values at}  $O(v_{\o})$.  \vspace{-0.5cm}
\item
{\em  The boost produces a dipole from a Doppler + aberrated modulation of the relative velocity,  i.e. $-2\big(\hsa v^a +\sigma_{\rm m}^{ab}\,v_b\big)_{\o}$ at  $O(v_{\o})$.}
\vspace{-0.2cm}
\item
{\em In perturbed FLRW, the aberrated dipole contribution $-2\big(\sigma_{\rm m}^{ab}\,v_b\big)_{\o}$ is second order in perturbations and would therefore be neglected at first order.}
\vspace{-0.2cm}
\item
{\em There is no boosted octupole or higher multipole. Any $\ell>2$ multipoles suggest observational systematics or a deficiency in theoretical modelling.
}\vspace{-0.2cm}
\end{itemize} 
}}\\

\noindent An example of an observational systematic that can induce $\ell>2$ multipoles and modify the $\ell\leq2$ multipoles (including the generation of a spurious dipole) is discussed in \autoref{sec3.2}.

The Doppler part of the dipole at leading order, $-2\hsa\,v^a$, is directed opposite to the relative velocity, since the observer's motion in the direction of receding galaxies reduces the effect of volume expansion in this direction. The aberration contribution $-2 \sigma_{\rm m}^{ab}\,{v}_b$ can modify this direction, especially if the shear strong.
The fact that there are no multipoles beyond the quadrupole is obvious from \eqref{14t}, since there are only 2 direction vectors $\tn^a$ involved in the boosted frame. This is
intuitively reasonable, since relative motion of the observer at the event o cannot induce intrinsic higher-order anisotropy. 

Equations \eqref{thhm}--\eqref{thhq} also reveal a critical property.
A heliocentric observer  measures
$\thsa$, $\tilde\hs^a$ and $\tilde\hs^{ab}$,
which together contain $1+3+5$ degrees of freedom. They are determined by the 9 degrees of freedom in $(\Theta_{\rm m}, v^a, \sigma_{\rm m}^{ab})_{\o}$. The system can be inverted numerically to find the matter expansion tensor and relative velocity at event o in terms of the measured multipoles.  
We can illustrate this via the approximate solution up to $O(v_{\o})$:
\begin{align}
 \Theta_{\rm m} &\circeq 3\thsa   +O(v^2)\,,
 \label{3.8}\\
\sigma^{ab}_{\rm m}  &\circeq    \tilde \hs^{ab}   +O(v^2)\,,\\
 v^a & \circeq -\frac{1}{2\thsa^2}\Big( \thsa\,\tho^a - \tho^{ab}\,\tho_b\Big) +O(v^2)\,.
 \label{3.10}
\end{align}
In summary: 

\fbox{\parbox{0.86\textwidth}{
\begin{itemize} \vspace*{-0.2cm}
\item
{\em the multipoles measured by a heliocentric observer can determine separately the average expansion rate,  the intrinsic shear and the relative velocity; 
 \vspace*{-0.2cm}
\item 
{\bf no} peculiar velocity information is needed.}  \vspace*{-0.2cm}
\end{itemize}}} \\
 
\noindent 
The case of perturbed FLRW is considered in \autoref{sec2.2}, where we provide more details on peculiar velocities. The independent measurement of the observer's relative velocity and the shear anisotropy can be seen as a covariant generalisation of the result found in \cite{Nadolny:2021hti} for the kinematic dipole in perturbed FLRW.
The practical implementation of the extraction of model-independent information about the kinematic state of the observer is presented in the follow-up paper \cite{Kalbouneh:2024tfw}.
There it is shown that the estimation of the observer’s velocity can be obtained in a completely model-independent manner, without making any assumptions about the metric, nor requiring information about the peculiar velocity field of galaxies, the background cosmology or the power spectrum of matter perturbations.

\subsection{FLRW and perturbed FLRW models}
\label{sec2.2}

The simplest example of \eqref{bhpobs}--\eqref{thhq} is provided by FLRW  spacetime, in which $\sigma_{\rm m}^{ab}=0$:
\bea \label{exfld}
\mbox{FLRW:}\quad \hs(n) \circeq \hsa \circeq H\quad \mbox{and}\quad
\tho(\tvn) \circeq \bigg[\gamma^2\Big( 1 +\frac{v^2}{3}\Big)-2\gamma v_a\,\tn^a + v_{\langle a}v_{b\rangle}\tn^a \tn^b\bigg]H\,.~~~
\eea
In this case, the physical Hubble parameter has only a monopole, while the boosted parameter has a dipole and quadrupole that are generated purely by the relative velocity.
The boosted FLRW observer sees an increased Hubble monopole, $H_0+O(v^2_{0})$, a dipole $-2H_0\,v_0^a+O(v^2_{0})$, which is directed opposite to the direction of relative motion, and an $O(v^2_{0})$ quadrupole $H_0\,v_{0}^{\langle a}v_{0}^{b\rangle}$.  
The next simplest examples, Bianchi I  and Lema\^itre-Tolman-Bondi spacetimes, are presented in Appendix~\ref{S:LTB}.  

The more realistic case is a linearly perturbed FLRW model. 
In a perturbative analysis, {\em galaxies move relative to their free-fall trajectories in the background spacetime, giving rise to so-called `peculiar' velocities}. In other words, peculiar velocities are measured relative to the background frame. Since this frame is gauge-dependent, it follows that
{\em peculiar velocities of matter sources are gauge-dependent}.  For example, peculiar velocities vanish in a comoving  gauge.
By contrast, in a general covariant analysis there is no background spacetime and consequently there are {\em no} peculiar velocities of galaxies. 
 
\begin{figure}[!t]
\centering
\includegraphics[width=0.6\textwidth]{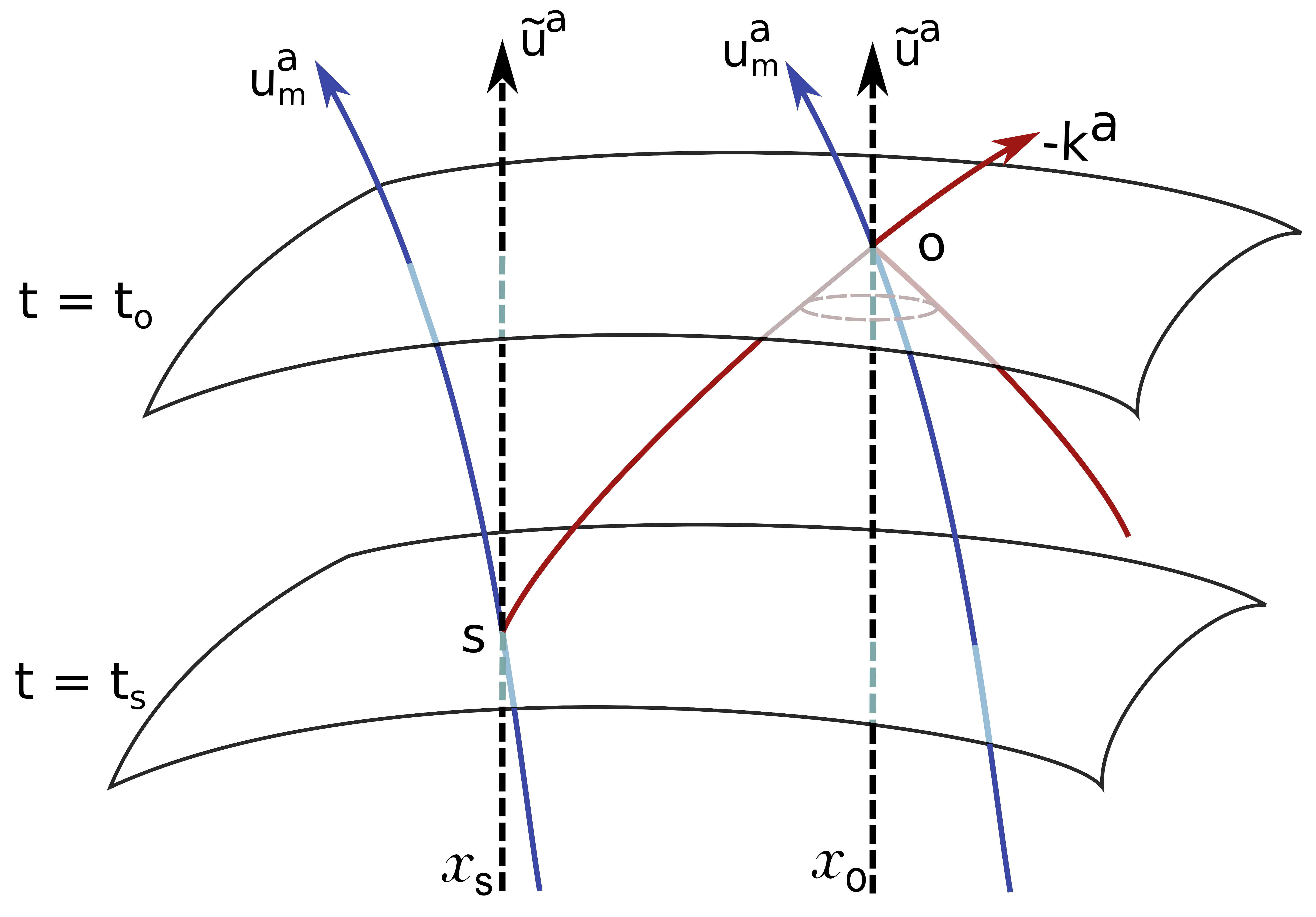} \\
\caption{Perturbed FLRW. 
}
 \label{fig0a}
\end{figure}

The covariant analysis in a general spacetime does not assume the Cosmological Principle, and only a single boosted observer at event o is required. By contrast, the perturbed FLRW model obeys the Cosmological Principle and therefore moving observers at each point on the  3-surface of constant cosmic time $t=t_0$ are statistically equivalent. At each cosmic time there is  a family of `gauge observers', with 4-velocity $\tu^\mu$, who are at fixed spatial coordinates $x^i$, i.e. at rest in the gauge frame (see \autoref{fig0a}).  
The gauge-frame observers see the matter as moving with relative velocity equal to the matter peculiar velocity, which is  considered as a first-order perturbed quantity. 
From a covariant viewpoint,  {\em perturbative peculiar velocities correspond to the velocities of boosted observers relative to the matter}.

In the Newtonian (or longitudinal) gauge, the metric at first order is
\bea \label{ds2ng}
\ud s^2=-\big(1+2\Phi \big)\ud t^2+a^2 \big(1-2\Phi \big)\ud \bm x^2\,,
\eea
where the background coordinates are $x^\mu=(t,x^i)$ and $\Phi$ is the  gravitational potential.
The 4-velocities of gauge observers and matter are (see \autoref{fig0a})
\bea \label{ng4v}
\tu^\mu=\big(1-\Phi,\bm 0 \big)\,,\quad 
u_{\rm m}^\mu=\Big(1-\Phi,-v^i  
\Big)~~ \mbox{with} ~~v^i=-\frac{\ud x^i}{\ud t}\,.
\eea
Here we defined the velocity $v^i$ of the gauge observers relative to the matter, in accordance with the convention chosen in the covariant relation \eqref{6}: 
$v^\mu=\tu^\mu-u^\mu_{\rm m}=(0,\,v^i)$.
As dictated by the Cosmological Principle, the gauge observer 4-velocity $\tu^\mu$ and the matter 4-velocity $u_{\rm m}^\mu$ both reduce to $\bar{u}^\mu=\delta^\mu_0$ in the background. This background 4-velocity represents the average 4-velocities of matter and the CMB,  in accordance with \eqref{cosprin}. 

Although gauge observers are at rest in the gauge frame, they are in fact accelerating: $\tu^\nu\nabla_\nu \tu^\mu=a^{-2}(0,\,\partial^i\Phi)$. By contrast, the
matter 4-acceleration vanishes: $u_{\rm m}^\nu\nabla_\nu u_{\rm m}^\mu=0$, as dictated by momentum conservation \eqref{momcg} for dust, which implies $\dot{v}^i+Hv^i=a^{-2}\partial^i\Phi$ (an overdot here is a derivative with respect to $t$). 
The projection tensor into the matter rest-space and the matter expansion tensor are 
\bea
h^{\rm m}_{\mu\nu} &=&a^2\begin{bmatrix}
0 & v_j \\
\\
v_i & \;\;\;  \big(1-2\Phi\big)\delta_{ij}
\end{bmatrix} , \\
\Theta^{\rm m}_{\mu\nu} &=&a^2\begin{bmatrix}
0 & Hv_j  \\
\\
Hv_i   & \;\;\;  \delta_{ij}\big( H-3H\Phi  +\dot \Phi\big)-\partial_i v_j
\end{bmatrix}  .
\label{thetaab}
\eea
Vanishing vorticity implies that
$\partial_{i} v_{j}$ in \eqref{thetaab} is symmetric.

From these equations we find the volume expansion and shear rates for matter:
\bea
\Theta_{\rm m} &=& 3 H+\delta\Theta_{\rm m} =
3H\Big(1-\frac{1}{3H}\partial_i v^i -\frac{\dot\Phi}{H}-\Phi \Big)\approx 
3H-\partial_i v^i, \label{pthet}\\
\sigma^{\rm m}_{ij}&=& -a^2 \partial_{\langle i} v_{j\rangle} ~~=-a^2 \Big( \partial_{i} v_{j}- \frac{1}{3}\partial_k v^k \delta_{ij}\Big)\,, \quad
\sigma^{\rm m}_{0\mu}=0\,.
\label{pshr}
\eea
In the second equality of \eqref{pthet}, the gravitational potential has been neglected relative to the velocity divergence: $|\Phi|\sim|\dot\Phi|/H \ll |\bm{\nabla}^2\Phi|/H^2\sim |\delta_{\rm m}| \sim |\partial_i v^i|/H$.
The covariant expression for the Hubble rate measured in the matter frame is \eqref{14xx}. In perturbed FLRW, \eqref{pthet} and \eqref{pshr} show that 
\begin{align}
\hs(n)=\Big(H-\frac{1}{3}\partial_iv^i\Big)
-{a^2} \partial_{\langle i} v_{j\rangle}\,n^i\,n^j\,.
\end{align}
The gauge observer measures a boosted Hubble parameter given by \eqref{14t}:
\bea \label{pertth}
\tho(\tvn)&=& {\Theta}^{\rm m}_{\mu\nu}\, \tilde K^\mu \tilde K^\nu  \notag\\
&=& -2\Theta^{\rm m}_{0i}\,\tu^0\, \tn^i +\Theta^{\rm m}_{ij}\,\tn^i \tn^j \quad \mbox{where}\quad \tn^\mu=(0,\tn^i) \notag\\
&=& 
\Big(H-\frac{1}{3}\partial_iv^i\Big)
-2H v_i\,\tn^i -{a^2} \partial_{\langle i} v_{j\rangle}\,\tn^i \tn^j \,,
\eea
in agreement with \eqref{bhpobs}--\eqref{thhq} at first order in perturbations.
{\em Since we used a covariant definition \eqref{14t} of the boosted Hubble parameter, the expression \eqref{pertth} is gauge invariant.}
We introduce the linear growth rate $f$ of large-scale structure formation:  $\dot{\delta}_{\rm m}=f H \delta_{\rm m}= \partial_iv^i$ (the second equality follows from energy conservation). Using the Poisson equation, we find  
\bea\label{evp}
v^i = \frac{2f}{3\Omega_{\rm m}{a^2}H}\,\partial^i\Phi\,.
\eea
Then  \eqref{pertth} becomes
\bea \label{pflbh}
\boxed{\tho = H \bigg[\Big(1-\frac{1}{3}f\delta_{\rm m}\Big) -\Big(\frac{4f}{3\Omega_{\rm m} {a^2}H} \,\partial_i\Phi\Big)\tn^i -\Big( \frac{2f}{3\Omega_{\rm m}H^2} \, \partial_{\langle i} \partial_{j\rangle} \Phi\Big)\tn^i \tn^j \bigg].}
\eea
In the perturbed FLRW model, where the spacetime geometry is assumed known and we use the field equations, \eqref{pflbh} shows that 
the boosted Hubble monopole can be expressed in terms of the growth rate of the matter density contrast $\delta_{\rm m}$, while the dipole and quadrupole are given by gradients of the gravitational potential.

The standard approach to fixing the value of the FLRW Hubble parameter $H$ using SNIa data is to assume a background cosmology and subtract from the redshift the perturbative contributions of peculiar velocities \eqref{evp}, estimated from all-sky galaxy distribution catalogs. 
Instead of this approach,
\eqref{pflbh}  provides an orthogonal, independent approach, which avoids the need for peculiar velocities.

We return now to the general, covariant case.

\section{Relating the Hubble parameter to cosmic distances}
\label{sec4}

In practice, measurement of the Hubble parameter is mainly via cosmic distances.
The covariant Hubble parameter $\hoo$ is not directly observable: in order to measure it, we need first to find its connection to cosmic distances. This requires a generalisation of the standard FLRW relations, without assuming a specific metric or using perturbative methods. We break the argument down to a series of steps as follows. \\

\noindent $\bullet$ We start with the relation between the matter-frame redshift $z$ and the lightray affine parameter $\lambda$, where $k^a=\ud x^a/ \ud \lambda$:
\begin{align} \label{dzdaf}
\frac{\ud z}{\ud \lambda} &= k^a\nabla_a(1+z)  
=\frac{1}{(u_{\rm m}^ck_c)_{\o}}\, k_ak_b\,\nabla^a u_{\rm m}^b 
= E_{\o} (1+z)^2\, \hs \,.
\end{align}
Here we used \eqref{ukin}, \eqref{15}, \eqref{14} and $k^b\nabla_b k^a=0$. It follows from \eqref{dzdaf} that the Hubble parameter at event o {$(z=0)$} is given in principle by
\bea \label{dzdafo}
\boxed{\hoo = \frac{1}{E_{\o}}\,\frac{\ud z}{\ud\lambda} \bigg|_{\o}\quad \mbox{in the matter frame}\,.}
\eea
Now we use the covariant boost relations
\bea \label{bhezl}
\boxed{\thoo=\Gamma_{\o}^{-2}\hoo\,,~~~\tilde E_{\o}=\Gamma_{\o}E_{\o}\,,~~~ 1+\tilde z= \Gamma_{\o}^{-1}(1+z)~~\mbox{and}~~ \tilde\lambda=\lambda\,,}
\eea
where the first one 
follows after using \eqref{tkkb} to rewrite the boost \eqref{14t} in terms of the unboosted $n^a$, i.e.,
 $\tho(\tn) \circeq \Gamma(n)^{-2} \,\hs(n)$.
Then
 the boost transformation of \eqref{dzdafo} is
\be
\boxed{\thoo = \frac{1}{\tilde E_{\o}}\,\frac{\ud \tilde z }{\ud \tilde\lambda}\bigg|_{\o}
\,,}
\ee
which shows that \eqref{dzdafo} {\em is invariant under boosts}. The boosted observer can use the same form of relation as \eqref{dzdafo}, simply replacing each quantity by its boosted counterpart.
\\

\noindent $\bullet$ Although $\lambda$ is a covariant and observer-independent variable, it is not observable and we need to convert it to an observable distance.
Since $\lambda$ has dimension length/energy (with $c=1$), $E_{\o}\lambda$ is a distance measure along past lightrays reaching the matter-frame observer at event o (with $\lambda_{\o}=0$). 
{\em In Minkowski spacetime, $E_{\o}\lambda$ corresponds precisely to the proper distance $d_P$ travelled by a lightray that reaches an inertial observer.} This is shown as follows:
\bea \label{kmumin}
\mbox{Minkowski:}~~
k^\mu &=&\frac{\Delta x^\mu}{\Delta \lambda}=
\frac{1}{\lambda-\lambda_{\o}}\big(-d_P,\,d_P \, n^i\big)
=\frac{d_P}{\lambda}\big(-1,\,n^i\big)\,,
\eea
where $x^\mu$ are inertial coordinates. By \eqref{12}, we also have $k^\mu = E_{\o}\big(-1,\,n^i\big)$ and therefore
\bea \label{kmumin2}
{d_P= E_{\o}\lambda\quad \mbox{in Minkowski}\,.}
\eea

\noindent $\bullet$ 
In a general spacetime, the Equivalence Principle implies that {\em any cosmic distance measure $d$ defined  via lightrays arriving at event {\rm o} and received by a matter observer, must reduce to the Minkowski distance $d_P$ close enough to event} o. It follows from \eqref{kmumin2} that for any cosmic distance measure $d$,
\bea \label{dalph}
\boxed{d = E_{\o}\lambda + O(\lambda^2)\quad \mbox{in the matter frame.}}
\eea
Differences between various distance measures appear at $O(\lambda^2)$.
Then by \eqref{dzdafo} we have
\bea \label{dddz}
{\hoo = \frac{\ud z}{\ud d} \bigg|_{\o}\quad \mbox{in the matter frame.}}
\eea
This is a covariant relation for any covariantly defined distance $d$,  since all quantities are given in the matter frame. Is \eqref{dddz} invariant under boosts?
As we show below,  under a boost of the observer, the relation \eqref{dddz} itself is {\em not} invariant for all possible distances $d$.

\subsection{Covariant cosmic distances}
\label{sec4.1d}

Distances in a general spacetime were defined covariantly by \cite{Kristian:1966} and then further clarified by \cite{Ellis:1971pg,MacCallum:1970} (see also \cite{Ellis:2012,Ivanov:2018cyw}). There are two distances which are of most relevance in cosmology: the area and luminosity distances.\\

\begin{figure}[t]
\centering
\includegraphics[width=0.6\textwidth]{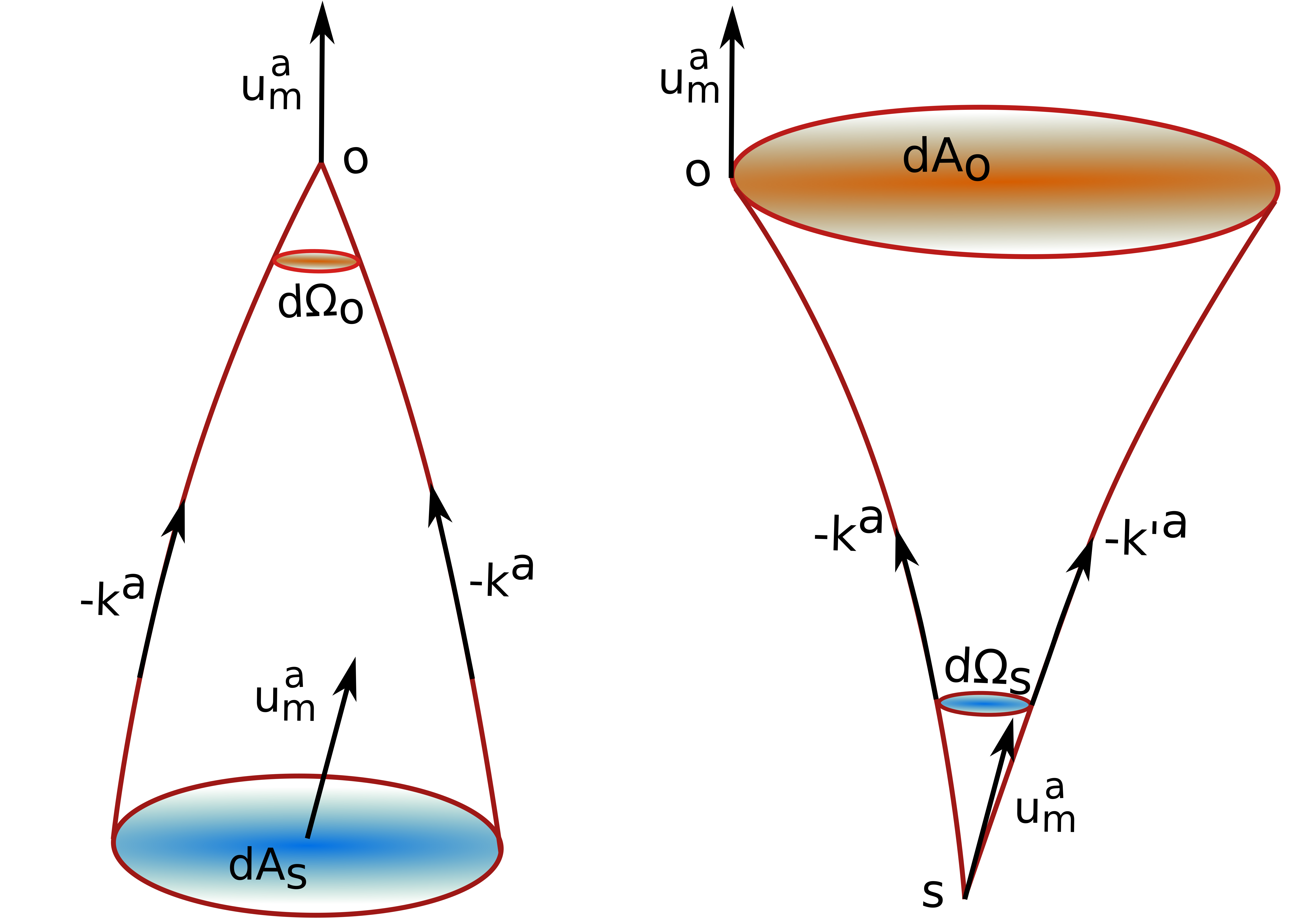} \\
\caption{Schematic to illustrate the definition of area distance $d_A$ (left) and luminosity distance $d_L$ (right). On the left, all lightrays  are on the past lightcone of event o. On the right, 
all lightrays  are on the future lightcone of event s, with
 one lightray  on the past lightcone of o. }
 \label{fig1}
\end{figure}

\noindent $\bullet$ The observer area distance $d_A$ is defined by
\bea \label{daad}
\boxed{\d A_{\e}= d_A^2\, \d\Omega_{\o}\quad \mbox{in the matter frame,}} 
\eea
where $\d A_{\e}$ is the area at the  source of the lightray bundle that converges at the  observer, while $\d \Omega_{\o}$ is the solid angle subtended by the source at the observer (see \autoref{fig1}).  The source and the observer are moving with the matter. 
The observer area distance $d_A$ generalises the familiar angular diameter distance to any spacetime. It is observable if there are `standard rulers' (e.g. the BAO feature) to estimate $\d A_{\e}$ for a measured $\d\Omega_{\o}$.

A boost $u^a_{\rm m}\big|_{\o}\to \tu^a_{\o}$ does not affect the intrinsic source area $\d A_{\e}$, but it does change the observer solid angle $\d\Omega_{\o}$:
\begin{align}
 \d \tilde A_{\e} = \d A_{\e}\,,~~\d\tilde\Omega_{\o}=\Gamma^{-2}_{\o}\,\d\Omega_{\o} \,.  
\end{align}
Then the covariant definition  \eqref{daad} leads to the  transformation of $d_A$ under a boost \cite{Maartens:2017qoa}:
\bea \label{tda}
\boxed{\tilde d_A(\tilde z, \tvn) = \Gamma_{\o}(n) \, d_A(z,n) 
=\gamma_{\o} \big(1+v_an^a \big)_{\o}\, d_A(z,n)\,.}
\eea

~\\ \noindent $\bullet$
The luminosity distance $d_L$ is defined by the ratio of fluxes at the source and at the observer:
\bea \label{dlle}
\boxed{d_L^2 = \frac{L_{\e}/4\pi}{F_{\o}} \quad \mbox{in the matter frame.}}
\eea
The lightray bundle is emitted from a point on the source and reaches the
observer with area  $\d A_{\o}$ on the screen in the observer's rest-space. 
This area is independent of the observer's 4-velocity (the `shadow theorem'  \cite{Sachs:1961zz,Ellis:1971pg}). The source luminosity  is also independent of the boost at o. However, the measured flux through the area at the observer is affected by boost transformations to (a)~the photon energy and (b)~the rate of reception of photons. Therefore
\begin{align}
\d \tilde A_{\o} = \d A_{\o}\,,~~\tilde L_{\e}=L_{\e}\,,~~  \tilde F_{\o} =\Gamma^2_{\o}\, F_{\o} 
\,.   
\end{align}
Then the covariant definition \eqref{dlle} leads to \cite{Maartens:2017qoa}:
\be \label{tdl}
\boxed{\tilde d_L(\tilde z,\tvn) = \Gamma_{\o}(n)^{-1} \, d_L(z,n) =\gamma^{-1}_{\o}\big(1+v_an^a \big)^{-1}_{\o}\, d_L(z,n)\,.}
\ee
In perturbed FLRW models, \eqref{tdl} recovers the relations in \cite{Hui:2005nm,Davis:2010jq}.\\

\noindent Equation \eqref{tdl} is consistent with \eqref{tda} by the reciprocity relation \cite{Ellis:1971pg,Ellis:2012}, 
\bea \label{recip}
d_L=(1+z)^2\,d_A\,,
\eea
which holds in any spacetime where photons are conserved. The covariant definitions of distances implicitly  assume the reciprocity relation. \\

\noindent $\bullet$ 
For a matter-frame observer, the Hubble parameter is determined by \eqref{dddz}:
\be \label{zdist}
\boxed{\hoo = \frac{\ud z }{\ud d}\bigg|_{\o} \quad \mbox{where}~~d=d_A~\mbox{or}~d_L~~\mbox{(matter frame).}}
\ee
This relation holds in the matter frame for any covariant distance measure -- but unlike \eqref{dzdafo}, it is {\em  not invariant under a boost for all distance measures} $d$, as we show in the following subsection. 

\subsection{Boosted observers need to use a corrected luminosity distance}
\label{sec3.1}

Measurements of $\hoo$ that use the redshift-distance  relation are made in the heliocentric frame. It is therefore important to investigate the behaviour of this relation under a boost.
The transformation of $\hoo$ and $z$ is given by \eqref{bhezl}, while  for $d$ we use \eqref{tda} and \eqref{tdl} to write
\begin{align}
\label{tdal}
\tilde d = \Gamma_{\o}^\alpha\, d \quad \mbox{where} \quad   \alpha=(1,-1)~~\mbox{for} ~~ d=\big(d_A\,, d_L\big)\,.
\end{align}
Then it follows that the boosted version of the redshift-distance relation \eqref{zdist} is given by
\begin{align}
\frac{\ud \tilde z }{\ud \tilde d}\bigg|_{\o}=
\Gamma_{\o}^{-1-\alpha}\,  \frac{\ud z }{\ud d}\bigg|_{\o}=\Gamma_{\o}^{-1-\alpha}\,\Gamma_{\o}^2\, \thoo
\quad\Rightarrow \quad \thoo = \Gamma_{\o}^{\alpha-1}\,
\frac{\ud \tilde z }{\ud \tilde d}\bigg|_{\o}\,.
\end{align}
In other words,  \eqref{zdist} {\em is preserved under a boost for the area distance, but \underline{not} for the luminosity distance:}
\be \label{zdist2}
\boxed{\thoo = \frac{\ud \tilde z }{\ud \tilde d_A}\bigg|_{\o} =\Gamma_{\o}^{-2}\, \frac{\ud \tilde z }{\ud \tilde d_L}\bigg|_{\o}.}
\ee

An observer  moving relative to the matter should use the modified redshift-luminosity distance relation  \eqref{zdist2}. Alternatively, if the standard relation  \eqref{zdist} is used in the moving frame, then the luminosity distance must first be corrected by an appropriate redshift factor. 
By \eqref{15y}, \eqref{tdl} and \eqref{zdist2}, we should use $d_{L*}=(1+z)^{-2} d_L$, which obeys the required transformation $\tilde{d}_{L*}= \Gamma_{\o}\,d_{L*}$:
\be \label{zdlum2}
\boxed{\hoo = \frac{\ud z }{\ud d_{L*}}\bigg|_{\o} \quad \mbox{implies}~~\thoo = \frac{\ud \tilde z }{\ud \tilde d_{L*}}\bigg|_{\o} \quad
\mbox{where}~~ d_{L*}=\frac{d_L}{(1+z)^2}.}
\ee
If photons are conserved, then $d_{L*}=d_A$ by the reciprocity relation \eqref{recip}. But $d_{L*}$ is an {\em independent way of measuring distance}.

The boost behaviour of $\hs_{\o}$ and $z$ enforces that \eqref{zdist} is boost-invariant only for a distance $d$ that obeys $\tilde d=\Gamma_{\o} d$.
We note that this holds for $d_A$, which is defined purely on the past lightcone of the  observer (it is purely geometric), while $d_L$ is defined using the future lightcone of the source and is not purely geometric \cite{Hasse:1999} (see \autoref{fig1}).

Suppose that the moving observer uses the unmodified luminosity distance, and thereby derives the {\em incorrect} boosted Hubble parameter 
\bea \label{inbhp}
\tilde{\hs}_{\o}^{\times}(\tvn)= \frac{\ud \tilde z }{\ud \tilde d_L}(\tvn)\bigg|_{\o}
=\thoo(\tvn) \,\tilde\Gamma_{\o}(\tvn)^{-2}
\,,
\eea
where $\thoo(\tvn)$ is the true boosted parameter, given by
\eqref{bhpobs} and $\tilde\Gamma_{\o}(\tvn)$ is given by \eqref{elbi}. 
What are the consequences of this error?
The incorrect Hubble parameter \eqref{inbhp} can be rewritten to leading order in velocity as
\bea \label{intho}
\tilde{\hs}^{\times}(\tvn)&\circeq & 
 \Big[\hsa -2\Big(\hsa\, v^a+ \sigma_{\rm m}^{ab} {v}_b\Big) \tn_a +\sigma_{\rm m}^{ab}\, \tn_a \tn_b\Big]\big(1-2\tilde{v}^c\tn_c \big)+ O(v^2)
\notag\\
&\circeq & \hsa  -\frac{6}{5}  \sigma_{\rm m}^{ab}\, {v}_b\, \tn_a +\sigma_{\rm m}^{ab}\, \tn_a \tn_b
+ 2 {v}^{\langle a} \,\sigma_{\rm m}^{bc\rangle}\, \tn_a \tn_b \tn_c + O(v^2) \,.
\eea
Here we used \eqref{bhpobs}, together with $\tilde v^a \circeq -v^a+O(v^2)$ and $\tilde h^{ab} \circeq h_{\m}^{ab}+O(v^2)$. We also used  
\bea \label{nanbnc}
\tn^{\langle a}\tn^b \tn^{c\rangle} = \tn^a\tn^b\tn^c -\frac{1}{5}\Big(\tilde h^{ab}\tn^c+ \tilde h^{ca}\tn^b+ \tilde h^{bc}\tn^a\Big),
\eea
which follows from the identity \eqref{ntfp} for $\ell=3$. 

Comparing \eqref{intho} with the correct version \eqref{bhpobs}--\eqref{thhq}, we conclude the following.  \\ 

\fbox{\parbox{0.9\textwidth}{
{\em A heliocentric observer who mistakenly  uses the incorrect redshift-luminosity distance relation \eqref{inbhp}, predicts an incorrect dipole in the Hubble parameter:\\ \hspace*{0.5cm} ${\bullet}$ the dominant Doppler part of the correct dipole, $-2\hsa\,v^a$, {\bf disappears}; \\ \hspace*{0.5cm} ${\bullet}$ the aberration  dipole term $\sigma_{\rm m}^{ab} {v}_b$ has the wrong factor ($-\frac{6}{5}$ instead of $-2$).
\\
 \hspace*{0.5cm} $\bullet$ Furthermore,  a spurious octupole is predicted.} } }  \\ \vspace*{0.15cm} 

\noindent This result applies at leading order in $v_{\o}$. At higher order in velocity, spurious $\ell>3$ multipoles will also arise. The main problem -- the incorrect removal of the dominant dipole contribution -- has a serious implication for  linearly perturbed FLRW models, in which $\sigma_{\rm m}^{ab} {v}_b$ is second order and is thus neglected:\\

\fbox{\parbox{0.9\textwidth}{
{\em   In linearly perturbed FLRW, using the incorrect redshift-luminosity distance relation in a boosted frame leads to a false prediction of {\bf no} dipole.}}}  \\ \vspace*{0.2cm}

\subsection{Higher-order terms in the redshift-distance relation}
\label{sec3.2}

The extraction of the Hubble parameter from redshift-distance measurements requires a derivative evaluated at the observer, which is $z_{\o}=0$ in the matter frame and $\tilde z_{\o}=\Gamma_{\o}^{-1}-1$ in the boosted frame. This amounts to taking the leading term in a Taylor expansion $z=\hoo \,d+O(d^2)$.
In reality it is not possible to precisely extract the $O(d)$  derivative at the observer, i.e. exactly at $d=0$. In practice,
small $d$ data needs to be extrapolated to $d=0$ and hence
the estimate of $\hoo$ can be contaminated by terms of $O(d^2)$. This requires the determination of these higher-order terms, which was first done theoretically by \cite{Kristian:1966} and \cite{MacCallum:1970}. Here we outline the key steps.  

We start with the redshift in the matter frame $1+z=( u_{\rm m}^ak_a)/( u_{\rm m}^bk_b)_{\o}= E_{\o}^{-1} u_{\rm m}^ak_a$ and expand it in the affine parameter:
\begin{align}
1+z(\lambda)&=\frac{1}{E_{\o}} \Big[ \big(u_{\rm m}^ak_a\big)_{\o} +\frac{\ud}{\ud\lambda}\big(u_{\rm m}^ak_a\big)\Big|_{\o}+\cdots \Big] 
\notag\\
\Rightarrow z(\hat\lambda) &=
\big(K_aK_b\nabla^a u_{\rm m}^b \big)_{\o}\, \hat\lambda + \frac{1}{2}\big(K_aK_bK_c \nabla^a\nabla^b u_{\rm m}^c\big)_{\o}\, \hat\lambda^2 +\cdots,
\label{kszl}
\end{align}
where we used $\ud/\ud\lambda=k_a\nabla^a$, with  $k^a_{\o}=E_{\o}K^a_{\o}$ and $\hat\lambda=E_{\o}\lambda$ (note that $\hat\lambda$ is an affine parameter). Then we need the area distance $d_A$ in terms of affine parameter. The geodesic deviation equation for lightrays implies that \cite{Kristian:1966}
\begin{align}
d_A^2 (\hat\lambda) = \hat\lambda^2\Big[1-\frac{1}{6}\big(K_aK_b R^{ab} \big)_{\o} \hat\lambda^2 -\frac{1}{12}\big( K_aK_bK_c \nabla^c R^{ab}\big)_{\o}\hat\lambda^3+\cdots \Big],
\label{ksgd}
\end{align}
where $R^{ab}$ is the Ricci tensor.
This is a geometric relation that expresses the effect of curvature on lightrays. Equations \eqref{kszl} and \eqref{ksgd} then imply that 
the covariant expansion of $z$  is given in any dust spacetime, in the matter frame, by  
\begin{align}
z &= \big(K_aK_b\Theta_{\rm m}^{ab} \big)_{\o}\, d + \frac{1}{2}\big(K_aK_bK_c \nabla^a\Theta_{\rm m}^{bc}\big)_{\o}\, d^2
\notag\\
&~+
\frac{1}{6} \Big[K_aK_bK_cK_d\nabla^a\nabla^b\Theta_{\rm m}^{cd}+ \frac{1}{2}\big(K_aK_b\Theta_{\rm m}^{ab} \big) K_cK_d R^{cd} \Big]_{\o} d^3
\notag\\
&~+
\frac{1}{24} \Big[K_aK_bK_cK_dK_e\nabla^a\nabla^b \nabla^c\Theta_{\rm m}^{de}+ {2}\big(K_aK_bK_c \nabla^a\Theta_{\rm m}^{bc} \big) K_dK_e R^{de} 
\notag\\
&\qquad\quad~ +
\big(K_aK_b\Theta_{\rm m}^{ab} \big) K_cK_d K_e\nabla^c R^{de} \Big]_{\o} d^4+O(d^5)\,, 
\label{zexpd2}
\end{align}
where $d=d_A$, but we can also use $d=d_{L*}$. This extends to $O(d^4)$ the expression in \cite{Kristian:1966} [which is given up to $O(d^2)$] and \cite{MacCallum:1970,Heinesen:2020bej} [up to $O(d^3)$]. In \cite{MacCallum:1970}, the expansion of $d$ in redshift is given up to $O(z^2)$, and this is extended to $O(z^3)$ in \cite{Heinesen:2020bej}.

Comparing  \eqref{zexpd2} with the FLRW luminosity distance expanded in redshift (e.g. \cite{Visser:2003vq}), we can identify covariant generalisations of the deceleration, $q_0=-\ddot a_0/(aH^2)_0\to \qoo$, jerk, $j_0=\dddot a_0/(aH^3)_0\to \mathbb{J}_{\o}$, and curvature,  $\Omega_{k0}\to \mathbb{R}_{\o}$, parameters \cite{Maartens:1980,Clarkson:1999zq,Clarkson:2010uz,Umeh:2013,Heinesen:2020bej,Kalbouneh:2024tfw}:
\begin{align}
\big(K_aK_b\Theta_{\rm m}^{ab} \big)_{\o} &= \hoo\,,\\
\big(K_aK_bK_c\nabla^c\Theta_{\rm m}^{ab}\big)_{\o} &  =\big(\qoo+3 \big)\hoo^2\,,
\label{qoxo}\\
\big(K_aK_bK_cK_d\nabla^a\nabla^b\Theta_{\rm m}^{cd}\big)_{\o} &= \big( \mathbb{J}_{\o} +10\qoo +15\big)\hoo^3\,,\\
\big(K_a K_b R^{ab}\big)_{\o} &=- 2 \big(\mathbb{R}_{\o}-\qoo-1\big)\hoo^2\,.
\end{align}
Here we will only consider the $O(d^2)$ term 
\begin{align}\label{xomulti}
\xoo \equiv \big(K_aK_bK_c\nabla^c\Theta_{\rm m}^{ab}\big)_{\o} \circeq \xoa+\xo^an_a+ \xo^{\langle ab\rangle} n_an_b+ \xo^{\langle abc\rangle} n_an_bn_c \,,
\end{align}
which defines the covariant deceleration parameter $\qoo$ via \eqref{qoxo}. For our purposes we do not need to compute the multipoles of $\qoo=\xoo/\hoo^2-3$, which are considerably more complicated than those of $\xoo$ alone, due to the $\hoo^{-2}$ factor.  (See \cite{Kalbouneh:2024tfw,Clarkson:1999zq,Clarkson:2010uz,Umeh:2013,Heinesen:2020bej,Heinesen:2021azp} for multipolar expansions of $\qoo$ and higher-order terms.)

In Appendix~\ref{appe}, we derive \eqref{xeq} for $\xoo$, which leads to the multipoles
\begin{subequations}
\begin{empheq}[box=\fbox]{align}
\xoa &\circeq  -\frac{1}{3}\dot{\Theta}_{\rm m}+ \frac{2}{9} \Theta^2_{\rm m}+\frac{2}{3}\sigma^{\rm m}_{ab}\sigma_{\rm m}^{ab}
\qquad \, \circeq  -\dot{\hsa}+2 \hsa^2+ \frac{2}{3}\,\hs_{ab}\,\hs^{ab} \,,
\label{xopol0a}
\\
\xo^a &\circeq  
 \frac{1}{3}h_{\rm m}^{ab}\, \nabla_b \Theta_{\rm m}+ \frac{2}{5}h_{\rm m}^{ab}\,h_{\rm m}^{cd}\, \nabla_c\sigma_{db}^{\rm m}   
~~ \circeq h_{\rm m}^{ab}\, \nabla_b \hsa +  \frac{2}{5}h_{\rm m}^{ab}\,h_{\rm m}^{cd}\, \nabla_c \hs_{db}\,,
\label{xopol1a}
 \\
\xo^{ab} &\circeq  -\dot{\sigma}_{\rm m}^{\langle ab \rangle} +\frac{4}{3}\Theta_{\rm m} \sigma_{\rm m}^{ab}+2
\sigma_{{\rm m}\,c}^{\langle a}\, \sigma_{{\rm m}}^{b\rangle c}
~\circeq -\dot{\hs}^{\langle ab \rangle} +4\hsa\,\hs^{ab}+2\,\hs^{\langle a}_c\, \hs^{b\rangle c}\,,
\label{xopol2a}
\\ 
\xo^{abc} &\circeq   \nabla^{\langle a} \sigma_{\rm m}^{bc \rangle}
\qquad \qquad \qquad \qquad \qquad \circeq \nabla^{\langle a} \hs^{bc \rangle} \,.
\label{xopol3a}
\end{empheq}
\end{subequations}
The multipoles of $\xoo$ are generated by the multipoles of $\hoo$ and their time and covariant spatial derivatives.
The spatial derivatives lead to an {\em intrinsic} dipole (i.e. not from any relative motion) in $\xoo$, generated by a spatial gradient of the 
$\hoo$ monopole, $(h_{\rm m}^{ab}\nabla_b \hsa)_{\o}$, and by a spatial divergence of the $\hoo$ quadrupole, $(h_{\rm m}^{ab}h_{\rm m}^{cd} \nabla_c \hs_{db})_{\o}$. There is also an intrinsic octupole, generated by the tracefree covariant derivative of the $\hoo$ quadrupole, $(\nabla^{\langle a} \hs^{bc \rangle})_{\o}$. The monopole and quadrupole of $\xoo$ are made up of products and time derivatives of the monopole and quadrupole of $\hoo$.

Contamination from local structure at a small distance $d=\varepsilon$ is given  by
\begin{align}
\hoo[\varepsilon] = \frac{\ud z}{\ud d}\bigg|_\varepsilon =\hoo+\varepsilon\xoo + O(\varepsilon^2)\,. 
\end{align}
Then \eqref{14xx} and \eqref{xopol0a}--\eqref{xopol3a} show that\\

\fbox{\parbox{0.92\textwidth}{
{\em at leading order, contamination of the  Hubble parameter from local structure near the matter observer introduces spurious effects, $\hoo \to \hoo[\varepsilon]$:\\ 
\hspace*{0.2cm} ${\bullet}$ an {\bf intrinsic} dipole $\varepsilon\xoo^a$ from spatial gradient of expansion and divergence of shear;
\\ \hspace*{0.2cm} ${\bullet}$ 
an intinsic octupole $\varepsilon\xoo^{abc}$ from spatial derivatives of shear;\\
\hspace*{0.2cm} ${\bullet}$  expansion and shear  from $\varepsilon\xoa_{\o}$ and $\varepsilon\xoo^{ab}$ modify the $\hoo$ monopole and quadrupole. 
\vspace*{-0.2cm}}}}\\ 

\noindent 
A key effect  of local structure is the spurious dipole in the matter-frame Hubble parameter, since in principle there should be no dipole. This dipole could be significant if measurements are taken at small enough distance or redshift -- since it is generated by spatial derivatives on small scales.
In principle,  $\xoo$  can be measured in the matter frame from distance data via
\bea \label{xodzdd}
\boxed{\xoo =\frac{\ud^2 z}{\ud \mskip1mu d^2}\bigg|_{\o}
\quad \mbox{for}\quad d= d_A~~\mbox{or}~~ d_{L*}=\frac{d_L}{(1+z)^2}\,,}
\eea
which follows from \eqref{zexpd2}. In practice, measurements are made in the heliocentric frame and we need to take into account both of the boost transformations, $\hoo\to\thoo$ and $\xoo\to\txoo$.
A boosted observer sees a kinematic dipole  
\eqref{thhd} in $\thoo$ -- but there is also a kinematic  contribution to the dipolar contamination in the boosted $\txoo$, given at leading order in Appendix~\ref{appb3} by \eqref{txa}. The leading-order contaminated Hubble dipole in the heliocentric frame is then
\begin{align}
\tho[\varepsilon]^a 
& \circeq  
-\frac{2}{3}\big( \Theta_{\m}v^a+3\sigma_{\rm m}^{ab}\,{v}_b \big) 
\notag\\
&~~~+
\frac{\varepsilon}{15}\bigg[ 
5\,{\rm grad}^a\,\Theta_{\rm m}+6 \big({\rm div}\,\sigma_{\rm m}\big)^a
+ 5\Big(3 \dot{\Theta}_{\rm m} - 2 \Theta^2_{\rm m}-6 \sigma_{\rm m}^2 \Big)v^a
\notag\\
&\qquad\qquad
+12\Big(3\dot{\sigma}_{\rm m}^{\langle ab \rangle} - 4\Theta_{\rm m} \sigma_{\rm m}^{ab} - 6\sigma_{\rm m}^{c\langle a} \sigma_{{\rm m}c}^{b\rangle}
\Big) v_b\bigg] +O(v^2,\varepsilon^2)\,,
\end{align}
where $\sigma_{\rm m}^2\equiv \sigma_{\rm m}^{ab}\sigma_{{\rm m}{ab}} $ and grad and div are the covariant spatial derivatives appearing in \eqref{xopol1a}.
A systematic investigation requires the effect not only of the deceleration parameter, but also of the higher-order parameters $\mathbb{J}_{\o}$ and $\mathbb{R}_{\o}$ on Hubble measurements. In addition to the dipole, the other multipoles need to be simultaneously considered.  
This is implemented in the follow-up paper \cite{Kalbouneh:2024tfw},
which uses a perturbed  FLRW model to calculate the theoretical multipoles of $\hoo$, $\qoo$, $\mathbb{J}_{\o}$ and $\mathbb{R}_{\o}$ and shows how to  retrieve,  in a fully model-independent way,  their individual values in an accurate  way.

Finally, we point out that in the definition \eqref{zdist} of the Hubble parameter, we use the relation of the redshift as an explicit function of distance, so that $\hoo$ is defined as the slope of this relation at the observer, corresponding to $d=0$.  
In practice, this means that redshifts are measured in bins of constant distance.
It is more usual to consider the Hubble diagram as a distance-redshift relation, so that we measure distance in bins of constant redshift:
\begin{align}
 \label{ditoz}
\boxed{
d = \frac{1}{\hoo}z+O(z^2) ~~\Rightarrow~~
\frac{1}{\hoo} = \frac{\ud  d }{\ud  z}\bigg|_{\o} \quad \mbox{where}~~ d=d_A\,,~ d_{L*}\,,
}
\end{align}
and for the heliocentric observer
\begin{align}
 \label{tditoz}
\tilde d = \frac{1}{\thoo}\big(\tz-\tz_{\rm o}\big)+O(\tz^2) ~~\Rightarrow~~
\frac{1}{\thoo} = \frac{\ud  \tilde d }{\ud  \tz}\bigg|_{\o} \quad \mbox{where}~~ \tz_{\rm o}=\Gamma_{\rm o}^{-1}-1\,.
\end{align}
Taylor expanding quantities as a function of redshift is also more practical, since redshift measurements are virtually error-free when compared to distance measurements. In Appendix~\ref{appc}, we give the basic relations for the distance-redshift case, where $\hoo^{-1}$ is extracted, rather than $\hoo$.
Note that \eqref{ditoz} is also invariant under boosts under the same condition as previously: {\em when extracting the inverse Hubble parameter from the distance-redshift relation, a moving observer should use the modified luminosity distance or the area distance.}

\newpage
\section{Conclusion}
\label{sec5}

Motivated by the tension between low- and high-redshift measurements of the Hubble constant,
we developed a model-independent and non-perturbative analysis of the Hubble parameter in a general spacetime, that is relativistically covariant. We started by identifying the physical tensors that encode the expansion rate  independent of the observer. Then we derived the covariant version of Lorentz transformations of key observables under a change of observer (\autoref{sec2.3c}). 

We defined the physical Hubble parameter $\hoo$ in \eqref{14}  as the projection of the local expansion tensor of dust matter onto the past lightcone -- this physical Hubble parameter is what a matter-frame observer would measure and it has only a monopole and a quadrupole induced by the shear~\eqref{14xx}. A boosted  (heliocentric) observer measures approximately the same
monopole and quadrupole, but also a kinematic dipole,  \eqref{bhpobs}--\eqref{thhq}. No higher multipole moments arise, if we neglect contamination and other systematics. 
The dipole has Doppler ($\propto v^a$) and aberration ($\propto \sigma_{\rm m}^{ab}v_b$) contributions -- the latter is necessarily neglected in linearly perturbed FLRW models.
The dipole is therefore only aligned with the peculiar velocity when the shear is small relative to the average (monopole) expansion rate. In that case, the Hubble rate will be smaller in the direction of the peculiar velocity, and larger in the opposite direction.

Two key results of our covariant cosmography are as follows, for a heliocentric observer:
\begin{itemize} \vspace*{-0.2cm}
    \item 
The volume and shear expansion rates, together with the relative velocity, can in principle be measured independently.
\item 
No peculiar velocity information is needed. Peculiar velocities, which are very difficult to measure accurately and precisely,  do not enter into a covariant cosmographic approach.
\end{itemize}

We applied our general covariant results to the standard perturbed FLRW models (\autoref{sec2.2}),
clarifying the nature of peculiar velocities as the gauge-dependent velocities of gauge observers relative to the matter.
We derived the perturbative covariant (and therefore gauge invariant) Hubble parameter for a boosted observer~\eqref{pflbh}. When the geometry is assumed and the Einstein equations are used, the boosted monopole is determined by the local growth rate and overdensity of large-scale structure, while the dipole and monopole are given by gradients of the gravitational potential. 
In our covariant approach,  peculiar velocity information is not needed.

The Hubble parameter is not directly measurable, but is typically extracted from a redshift-distance relation. In order to generalise this approach in covariant cosmography, we first derived the covariant relation between cosmic distances, the affine parameter of lightrays and $\hoo$ \eqref{dalph}--\eqref{dddz}. Then we presented a covariant treatment of area and luminosity distances in a general spacetime, deriving their  transformation under boosts (\autoref{sec4.1d}).
For an observer moving with the matter, the physical Hubble rate coincides with the slope of the redshift-distance relation at the observer, \eqref{zdist}. This holds for any covariant distance measure, including area and luminosity distances. 

For a heliocentric observer however, the case is more subtle. {Provided that the area distance is used, the boosted Hubble parameter $\thoo$ remains equal to the slope of the boosted redshift-distance relation at the observer.} This is not the case for the luminosity distance, which requires a boost correction in order for $\thoo$ to match the boosted cosmographic redshift-distance relation. Using the uncorrected luminosity distance  gives a spurious correction to the measured dipole in the Hubble rate~-- as well as a spurious octupole. In a linearly perturbed FLRW model, this leads to a false prediction of {\em no dipole}.
Therefore, in order to use cosmographic distances measured by magnitudes or luminosities, the luminosity distance must be converted to $d_{L*}$ as in
\eqref{zdlum2}. This is equal to the area distance if photons are conserved, but it is an independent observable, directly related to the magnitude. 

These results assume that one can extract the slope of the  redshift-distance relation in a simple way. In practice, we cannot measure a derivative at the observer
and extrapolating to $d=0$ leads to contamination from $O(d^2)$ terms, induced by  local nonlinear structure.
The covariant corrections at $O(d^2)$ result in a generalised deceleration parameter, given by \eqref{xomulti}--\eqref{xopol3a} and  Appendices~\ref{appe}--\ref{appb3}. This generates spurious contributions to $\hoo$ from an intrinsic dipole and an intrinsic octupole -- i.e. not induced by relative motion, but by spatial derivatives. For a heliocentric observer, the kinematic dipole in the Hubble parameter $\thoo$ is contaminated (at leading order in $d$) by the intrinsic dipole and octupole in $\txoo$, in addition to further spurious kinematic contributions from the boost transformations (see Appendix \ref{appb3}). This implies that the region of the redshift-distance relation where the dipole and octupole are minimised is the region probing the true matter expansion rate. 

We showed explicitly how the observer's velocity can be extracted from the heliocentric Hubble parameter simultaneously with the volume and shear expansion rates -- and without peculiar velocity information.  The same method will apply when the deceleration parameter is treated together with Hubble parameter, but with the complication that the deceleration parameter has an intrinsic dipole in the matter frame and additional kinematic dipolar contributions in the heliocentric frame.

The multipoles of the Hubble and deceleration parameters
are physically defined in the matter frame. However, they are not measured in this frame, but in the heliocentric one. Therefore cosmographic estimates of these multipoles must apply the correct boost transformations -- not only to the redshift, but also to the multipoles themselves. In order to avoid a `theoretical'  systematic,  the heliocentric multipoles must be estimated by correctly factoring out the observer's motion. 

In this paper we are not dealing with practicalities of measurement.
An implementation of our covariant cosmographic approach is presented in the follow-up paper \cite{Kalbouneh:2024tfw}, which uses
a detailed analytical model that is physically motivated and consistent with current observations. This model builds on the earlier paper \cite{Kalbouneh:2022tfw}, which applied a cosmographic analysis to distance data from CosmicFlows-3 (galaxies) and Pantheon (supernovae). 
The follow-up paper \cite{Kalbouneh:2024tfw} also forecasts the precision possible from upcoming observations by the Zwicky Transient Facility.

We note that it is not optimal
to extract the Hubble parameter directly from distances \cite{Kalbouneh:2022tfw,Kalbouneh:2024tfw}. 
The optimal way is to use the distance modulus, i.e. the logarithm of distance, whose errors are Gaussian in the standard model. This approach is used by \cite{Kalbouneh:2022tfw,Kalbouneh:2024tfw} to define the covariant `expansion rate fluctuation field',
\begin{align}
    \eta(z,n)=\log \left[\frac{z}{d_L(z,n)}\right]- {\cal M}(z)\,,
\end{align}
where ${\cal M}$ is the monopole, i.e. the all-sky average of $\log(z/d_L)$.

Finally, we note that
there are deeper problems associated with local nonlinear structure around the observer. 
Although the covariant analysis is fully nonlinear, the assumption of a dust model for matter on small scales
could break down and lightray propagation could be affected. This also applies in the standard model where an FLRW background is assumed \cite{Umeh:2022hab,Umeh:2022prn,Umeh:2022kqs}.
Cold baryons and dark matter particles behave like dust before shell crossing occurs on small scales. In order to avoid the shell-crossing problem, a kinetic theory model of a collisionless gas can be used (e.g. \cite{1983AnPhy.150..455E, ETM1983,Buchert:1997dr, Lewis:2002nc,Bartelmann:2019unp,Fardeau:2023sxl}). However this does not address the next level of the problem: when  the fundamental particles form virialised halos and galaxies. These virialised objects then become the ‘dust particles’ in the standard paradigm. It is not clear how to treat this situation consistently (see e.g. \cite{Biagetti:2015hva,Matsubara:2019tyb}) -- and therefore it is not clear how to estimate the consequences for measurements made at $d=0$. It is conceivable that there are averaging  and backreaction effects, which could impact not only the matter model, but also lightray propagation and possibly the spacetime metric itself (e.g. \cite{Clarkson:2011zq,Adamek:2017mzb,Heinesen:2018vjp, Fanizza:2019pfp,Buchert:2022zaa,Wagner:2022etu,Anton:2023icm}). 

Regarding the Hubble tension, it seems possible that the model-dependent perturbed FLRW framework is inadequate to resolve the problem. The solution may require a resolution of these deeper problems of modelling matter, in combination with a model-independent covariant cosmography.

\vfill
\noindent{\bf Acknowledgements:} We thank Raul Abramo, George Ellis, Pedro Ferreira, Asta Heinesen, Obinna Umeh and Jenny Wagner for useful comments and helpful discussions. RM is supported by the South African Radio Astronomy Observatory and the National Research Foundation (grant number 75415). 
JS was supported by the Hellenic Foundation for Research \& Innovation (H.F.R.I.), under the `First call for H.F.R.I. research projects to support faculty members and researchers and the procurement of high-cost research equipment' Grant (Project No. 789)
and by the Taiwan National Science \& Technology Council. CC is supported by the UK Science \& Technology Facilities Council Consolidated Grant ST/T000341/1.
BK and CM are supported by the {\it Programme National Cosmologie et Galaxies} (PNCG) and {\it Programme National Gravitation R\'ef\'erences Astronomie M\'etrologie} (PNGRAM) of CNRS/INSU with INP and IN2P3, co-funded by CEA and CNES.

\newpage

\appendix

\section{Appendix}
\subsection{General conservation equations}
\label{appa}

The energy-momentum tensor  $T_{\rm f}^{ab}$ for a non-interacting fluid f in a general spacetime is given by
\begin{align}   \label{enmom}
 T_{\rm f}^{ab}&=\rho\, u^a u^b +P h^{ab} + q^a u^b +q^b u^a +\pi^{ab}\quad\mbox{where}~~q_au^a=0\,, ~ \pi_{ab}u^b=0=\pi_a^a\,,
\end{align}
 where $u^a$ is an arbitrarily chosen 4-velocity. 
The covariant energy conservation equation ($u_a\nabla_b T_{\rm f}^{ab}=0$) and momentum conservation equation ($h_c^a\nabla_b T_{\rm f}^{cb}=0$) are \cite{Ellis:1971pg,Ellis:2012}:
\bea 
\label{encg}
&& \dot\rho + (\rho+P)\Theta+h^{ab}\nabla_a q_b+ 2 \dot{u}^aq_a +\sigma^{ab}\pi_{ab}=0\,, 
\\ \label{momcg}
&& h^b_a \dot{q}_b +\frac{4}{3}\Theta q_a + (\rho+P) \dot{u}_a + h^b_a\nabla_b P+ h^d_a h^{bc}\nabla_c \pi_{bd} 
 +\sigma_{ab}q^b + \omega_{ab}q^b +\pi_{ab}\dot{u}^b =0\,,
\eea
where an overdot denotes $u^a\nabla_a$.
Recall that $T_{\rm f}^{ab}$ is the covariant physical quantity, while the dynamical ($\rho, P,\dots$) and kinematic ($\Theta, \dot u^a,\dots$) quantities depend on the choice of $u^a$.

For a dust fluid, $P_{\rm m}=0=\pi_{\rm m}^{ab}$ and $q_{\rm m}^a=0$ in the dust frame, defined by the 4-velocity $u_{\rm m}^a$. Then it follows from \eqref{momcg} that $\rho_{\rm m} \dot{u}_{\rm m}^a=0$. Consequently, \eqref{encg} and \eqref{momcg} lead to
\begin{align}
\dot{u}_{\rm m}^a=0\,,\quad
\dot{\rho}_{\rm m}+{\rho}_{\rm m}\Theta_{\rm m}=0 \,.
\end{align}
Note that a moving observer measures different dynamical quantities for a dust fluid \cite{Maartens:1998xg}:
\begin{align}
T_{\rm m}^{ab} &= \rho_{\rm m}\,u_{\rm m}^a\,u_{\rm m}^b=
\tilde\rho\,\tu^a\,\tu^b+
\tilde P \tilde h^{ab} + \tilde q^a \tu^b +\tilde q^b \tu^a +\tilde\pi^{ab}\\
\label{trho}
\Rightarrow\quad\tilde \rho &= \gamma^2 \rho_{\m}\,, ~~
\tilde P = \frac{1}{3}\gamma^2 v^2 \rho_{\m}\,,~~
\tilde q^a = \gamma^2\rho_{\m}\, \tilde v^a\,,  ~~
\tilde{\pi}^{ab} = \gamma^2\rho_{\m}\, \tilde v^{\langle a}\, \tilde v^{b\rangle}\,,
\end{align}
where $\tilde v^a$ is given by \eqref{6x}. The pressures and momentum flux measured by the observer who is moving relative to the matter frame are not physical. They arise purely from relative motion and all vanish if $v^a=0$.
\\

\subsection{Covariant multipole expansion}
\label{appcovm}

The covariant multipoles are tracefree tensors in the observer's rest space that generalise a spherical-harmonic decomposition. They are given in general by \cite{1983AnPhy.150..455E,Gebbie:1998fe,Maartens:1998xg,Challinor:1998aa}: 
\begin{align}\label{a7}
F=\sum_{\ell=0}^\infty F_{A_\ell}\,n^{\langle A_\ell\rangle}\quad\mbox{and}\quad
F_{A_\ell}=\frac{(2\ell+1)!}{4\pi(\ell!)^22^\ell}\,\int\ud \Omega_n\,F\,n_{\langle A_\ell\rangle} = F_{\langle A_\ell\rangle}\,,
\end{align}
where $A_\ell\equiv a_1\,a_2 \cdots a_\ell$ and
$n^{\langle A_\ell\rangle}\equiv n^{\langle a_1}\,n^{a_2} \cdots n^{a_\ell\rangle}$.
This formula recovers the result \eqref{14xx}, using
the identity
\begin{equation}\label{a8}
\int\ud \Omega_n\,n^{A_\ell} 
= \frac{4\pi}{\ell+1} 
\begin{cases}
  0 &   \ell~\text{odd} \\
 h_{\m}^{(a_1a_2} h_{\m}^{a_3 a_4}\cdots h_{\m}^{a_{\ell-1}a_\ell)} & \ell~\text{even}
\end{cases}
\end{equation}
Another useful identity is
\begin{align}\label{ntfp}
n^{\langle A_\ell\rangle} =\sum_{k=0}^{[\ell/2]} \Delta_{\ell k}\,h_{\m}^{( A_{2k}} \,n^{A_{\ell-2k})} \quad \mbox{where}\quad
\Delta_{\ell k} &= \frac{(-1)^k\, \ell!\, (2\ell-2k-1)!!}{(\ell-2k)!\, (2\ell-1)!!\, (2k)!!}\,.
\end{align}
Here $[\ell/2]=$ the largest integer part of $\ell/2$ and $m!! \equiv m(m-2)(m-4)\cdots 2$ or 1.

\subsection{Bianchi I LRS and LTB spacetimes}
\label{S:LTB}

The simplest example of the boosted and matter-frame Hubble parameters is FLRW spacetime with dust matter, given by \eqref{exfld}. If we relax the isotropy of FLRW, the next simplest model is the homogeneous Bianchi I LRS (locally rotationally symmetric) spacetime:
\begin{equation}
    \d s^2 = -\d t^2 + a_\|^2(t)\,\d x^2 + a_\perp^2(t)\big( \d y^2+\d z^2\big)\,.
\end{equation}
By homogeneity, any event on a 3-surface $t=t_0$ can be chosen as the observer event o. If we relax the homogeneity of FLRW, the next simplest example is the 
Lema\^itre-Tolman-Bondi (LTB) spacetime,
\begin{equation}
    \d s^2 = -\d t^2 + A_\|^2(t,r)\,\d r^2 + A_\perp^2(t,r)\, \d\Omega^2\,,
\end{equation}
which is isotropic about the worldline $r=0$. 
By isotropy, all angular positions on spheres of radius $r$ centred at $r=0$ are equivalent.
In both spacetimes, $x^0=t$ is proper time along dust worldlines, with $u_{\rm m}^\mu=\delta^\mu_0$. (See e.g. \cite{Ellis:2012} for further details.)

By symmetry, the matter shear vanishes at the LTB centre, $\sigma_{\rm m}^{ab}(t_0,r_{\o}=0)=0$. Therefore the matter and boosted observers at the LTB centre see the same structure of matter-frame and boosted Hubble parameters as in FLRW, i.e. as in \eqref{exfld}. 

Off-centre observers in LTB, i.e. at $(t_0,r_{\o}>0)$, see a matter shear that has the same form as in Bianchi I LRS. In both cases, there is a preferred direction $e^\mu$ for the observer, where $e_\mu u_{\rm m}^\mu=0$ and $e_\mu e^\mu=1$. In LTB, $e^\mu$ is the radial direction at  $(t_0,r_{\o}>0)$, away from the centre. In Bianchi I LRS, $e^\mu$ is the  direction of the spatial $x$-axis of local rotational symmetry.

In both spacetimes, there is an expansion rate  $H_\|$ along the $e^\mu$ direction and an expansion rate $H_\perp$ in the screen space orthogonal to $e^\mu$. 
The average expansion rate of matter worldlines is therefore
\begin{align} \label{bhsa}
\hsa =\frac{1}{3} \big(H_\|+2H_\perp\big) \,.
\end{align}
Screen space has the metric $S^{\mu\nu}=h_{\rm m}^{\mu\nu}-e^\mu e^\nu$
and the  matter shear must be proportional to $2 e^\mu e^\nu-S^{\mu\nu}$ (since shear is tracefree) and to $H_\|-H_\perp$ (since shear is generated by  anisotropic expansion). A simple calculation gives
\begin{align} \label{bsig}
\sigma_{\rm m}^{\mu\nu}=\frac{1}{3}\big(H_\|-H_\perp \big) \big(2 e^\mu e^\nu-S^{\mu\nu} \big) \,.  
\end{align}
Note that in LTB spacetime, $H_\|=H_\perp$ at $r_{\o}=0$.

It follows from \eqref{bhsa} and \eqref{bsig} that the covariant matter-frame Hubble parameter \eqref{14xx} in both spacetimes has the form
\begin{align}
{\hs(n) \circeq \frac{1}{3} \big(H_\|+2H_\perp\big) 
+ \big(H_\|-H_\perp \big) e_{\langle\mu} e_{\nu \rangle} n^\mu n^\nu \,.}
\end{align}
For the boosted observer, the general relations \eqref{bhpobs}--\eqref{thhq} describing the multipoles of $\tilde\hs(\tn)$ give the following in Bianchi I LRS and LTB spacetimes:
\begin{align}
\thsa &\circeq \gamma^2 \bigg[\frac{1}{3}\Big( 1+\frac{1}{3}v^2\Big)\big(H_\|+2H_\perp\big)+ \big(H_\|-H_\perp \big) e_{\langle\mu} e_{\nu \rangle} v^\mu v^\nu \bigg] \,,\\
\tilde\hs_\mu &\circeq -2\gamma \bigg[\frac{1}{3}\big(H_\|+2H_\perp\big)v_\mu+ \big(H_\|-H_\perp \big) e_{\langle\mu} e_{\nu \rangle}  v^\nu \bigg]\,,\\
\tilde\hs_{\mu\nu} &\circeq \big(H_\|-H_\perp \big) e_{\langle\mu} e_{\nu \rangle}  +\frac{1}{3}\big(H_\|+2H_\perp\big) v_{\langle\mu} v_{\nu \rangle}\,.
\end{align}

\subsection{Covariant deceleration parameter and its multipoles}
\label{appe}

We begin with the second-order term   in \eqref{zexpd2}:
\bea
\xo &=& K_c\nabla^c\big(\Theta_{\rm m}^{ab} K_aK_b \big) - \Theta_{\rm m}^{ab}\,K_c\nabla^c \big(K_aK_b \big) \notag \\
&=& K_c\nabla^c \hs-2\Theta_{\rm m}^{ab}\,K_a \big( -\hs \big) K_b 
=- \dot{\hs}+2\hs^2 +n_a\nabla^a \hs \,,\label{xeqn}
\eea
where we used $K_b\nabla^bK^a = - \hs K^a$. By \eqref{14xx} the last term on the right is
\bea \label{ndelh}
n_a\nabla^a \hs &=& \frac{1}{3}n_a\nabla^a \Theta_{\rm m}+n_an_b n_c \nabla^a\sigma_{\rm m}^{bc} +
2\sigma_{\rm m}^{bc}\,n_c n_a \nabla^a n_b \,.
\eea
The derivative of $n^a$ along its own direction  may be written as
\bea
 n_a \nabla^a n^b = n_a \nabla^a K^b + n_a \Theta_{\rm m}^{ab} = - \hs K^b +\dot{n}^b + n_a \Theta_{\rm m}^{ab}~~\mbox{where}~~ \dot{n}^b=u_{\rm m}^a\nabla_a n^b\,. \label{ndeln}
\eea
It appears that we need to determine 
 $\dot{n}^a$ at the  observer event o, but in fact this term is cancelled by a term in $\dot{\hs}$ in \eqref{xeqn}: $-(\sigma_{\rm m}^{ab}\,n_a n_b)^{\!\displaystyle\cdot}=
 -\dot{\sigma}_{\rm m}^{ab}\,n_a n_b- 2\sigma_{\rm m}^{ab}\,n_a\dot n_b$. In addition, the hexadecapole term $2\sigma_{\rm m}^{bc}\,n_c (-\sigma_{\rm m}^{de}\,n_d n_e\,  n_b)$, from  \eqref{ndelh} and \eqref{ndeln}, is cancelled by the opposite-sign shear squared term in $2\ho^2$ in \eqref{xeqn}. The absence of a hexadecapole is required by the definition of $\xo$ in \eqref{xomulti}.
Collecting the above relations we have
\begin{align} \label{xeq}
\xo &= -\frac{1}{3}\dot\Theta_{\m}-\dot\sigma_{\m}^{ab}\,n_an_b  +\frac{2}{9} \Theta_{\m}^2+ 
\frac{4}{3}\Theta_{\m}\, \sigma_{\m}^{ab}\,n_an_b 
\notag\\ &~~
+\frac{1}{3}n_a\nabla^a \Theta_{\m}
+2\, \sigma_{\m c}^{a}\,\sigma_{\m}^{bc} \,n_a n_b
+n_a n_b n_c \nabla^a \sigma_{\m}^{bc}\,.
\end{align}
Then we use the identity \eqref{nanbnc} for $n^an^bn^c$, which  
leads to the $\xoo$ multipoles 
\eqref{xopol0a}--\eqref{xopol3a}.

Contamination will also affect the boosted Hubble parameter $\thoo$ via $\txoo$. The boost transformation $\tilde K^a=\Gamma_{\o}^{-1} K^a$ implies that
\begin{align}
\txoo =\Gamma_{\o}^{-3}\, \xoo\,,    
\end{align}
and together with 
\eqref{15y},  \eqref{tda} and \eqref{tdl}, this leads to 
boost invariance of \eqref{xodzdd}:
\bea
\txoo =\frac{\ud^2 \tz}{\ud \mskip1mu \tilde d^2} \bigg|_{\o}\quad \mbox{for}\quad \tilde d= \tilde d_A~~\mbox{or}~~ \tilde d_{L*}=\frac{\tilde d_L}{(1+\tz)^2}\,.
\eea
The boost transformation of the multipoles \eqref{xopol0a}--\eqref{xopol3a} is given in Appendix~\ref{appb3}.

What is the relation of $\xoo$ to the standard deceleration parameter?
In an FLRW spacetime, \eqref{xopol0a} reduces to
\bea \label{fxoo}
\mbox{FLRW:} \qquad
\xoo = \xoa_{\o}= \big(-\dot H + 2H^2\big)_0 = (q_0+3)H_0^2\,,
\eea
where $q_0=-\ddot a_0/(aH^2)_0$ is the FLRW deceleration parameter.
This motivates the definition of the covariant deceleration parameter
\begin{align}
 \label{qoo}  
\qoo=\frac{\xoo}{\hoo^2}-3~~\Rightarrow~~ 
\qoo =\bigg[\Big(\frac{\ud z}{\ud \mskip1mu d} \Big)^{\!-2}\,\frac{\ud^2 z}{\ud \mskip1mu d^{\,2}}\bigg]_{\o}-3\,. 
\end{align}
It follows that under a boost:
\begin{align}
 \label{tqoo}  
\tqoo=\Gamma_{\o}{\qoo}+3\big(\Gamma_{\o}-1 \big)\,, 
\end{align}
where we used $\thoo=\Gamma_{\o}^{-2}\hoo,~\txoo=\Gamma_{\o}^{-3}\xoo$. Together with \eqref{qoo}, this implies that
\begin{align}
\label{tqoo2}
\bigg[\Big(\frac{\ud \tilde z}{\ud \mskip1mu \tilde d} \Big)^{-2}\frac{\ud^2 \tilde z}{\ud \mskip1mu \tilde d^2}\bigg]_{\o}
= \Gamma_{\o} \bigg[\Big(\frac{\ud z}{\ud \mskip1mu d} \Big)^{\!-2}\,\frac{\ud^2 z}{\ud \mskip1mu d^{\,2}}\bigg]_{\o}
\,,   
\end{align}
 for any cosmic distance $d$. This means that\\

\fbox{\parbox{0.9\textwidth}{
{\em 
 the redshift-distance relation in \eqref{qoo} is not invariant under a boost for any cosmic distance $d$. An observer moving relative to the matter must use the boosted redshift-distance relation \eqref{tqoo2}.}}}
 \vspace*{0.5cm}
 
\subsection{Boost transformation of deceleration parameter multipoles}
\label{appb3}

The boost transformation of $\qoo$ follows from the boost of $\xoo$, using \eqref{qoo} and \eqref{tqoo}. By \eqref{zexpd2}:
\begin{align}
 \txo(\tn)\circeq  \tilde K_a\, \tilde K_b\, \tilde K_c\, \nabla^c\Theta_{\rm m}^{ab}\,. 
\end{align}
This leads to a multipole expansion  as seen by the moving observer of the form
\bea \label{xopol2}
\txo &\circeq & \tilde{\xoa} + \tilde{\xo}_{a}\, \tilde{n}^a +\tilde{\xo}_{ab}\,\tilde{n}^{\langle a}\tilde{n}^{b\rangle}
+\tilde{\xo}_{abc}\,\tilde{n}^{\langle a}\tilde{n}^b\tilde{n}^{c\rangle}\,,
\eea
where the projections and tracefree parts  are defined in terms of $\tilde{h}^{ab}$ not $h_{\m}^{ab}$.  We find that to leading order in velocity:
\begin{align}
\tilde{\xoa} &\circeq \xoa  
-\frac{5}{3}\xo_a v^a +O(v^2)\,,\\
\tilde{\xo}_a  &\circeq 
\xo_a-3\xoa v_a -\frac{12}{5}\xo_{ab}v^b +O(v^2)\,,\\
\tilde{\xo}_{ab} &\circeq 
\xo_{ab} -2\xo_{\langle a}v_{b\rangle} - 3 \xo_{abc}v^c +O(v^2)\,,\\
\tilde{\xo}_{abc} &\circeq \xo_{abc}
-\xo_{\langle ab}v_{c \rangle} +O(v^2)
\,.
\end{align}
Here $\xoa$, $\xo_a$, $\xo_{ab}=\xo_{\langle ab \rangle}$ and $\xo_{abc}=\xo_{\langle abc \rangle}$ are given by \eqref{xopol0a}--\eqref{xopol3a}, leading to:
\begin{align}
\tilde{\xoa} &\circeq 
-\frac{1}{3}\dot{\Theta}_{\rm m}+ \frac{2}{9} \Theta^2_{\rm m} +{\frac{2}{3}}\sigma^{\rm m}_{ab}\sigma_{\rm m}^{ab}  - \frac{1}{9} \Big(5\, \nabla_a \Theta_{\rm m}
+ 6\,h_{\rm m}^{bc}\, \nabla_c\sigma_{ab}^{\rm m}\Big)v^a+O(v^2) \,, \\
\tilde{\xo}^a &\circeq  
\frac{1}{3}\, h_{\rm m}^{ab}\nabla_b \Theta_{\rm m}+ \frac{2}{5}\,h_{\rm m}^{ab}\,h_{\rm m}^{cd}\, \nabla_c\sigma_{db}^{\rm m}
+ \frac{1}{3}\Big(3\, \dot{\Theta}_{\rm m} - 2\, \Theta^2_{\rm m}
-6\, \sigma_{\rm m}^{bc}\sigma^{\rm m}_{bc} \Big)v^a
\nonumber\\&\;\;\;
+\frac{4}{5}\Big(3\,\dot{\sigma}_{\rm m}^{\langle ab \rangle} - 4\,\Theta_{\rm m} \sigma_{\rm m}^{ab} - 6\,\sigma_{\rm m}^{c\langle a} \sigma_{{\rm m}c}^{b\rangle}
\Big) v_b +O(v^2)\,,
\label{txa}\\
\tilde{\xo}^{ab}       &\circeq -\dot{\sigma}_{\rm m}^{\langle ab \rangle} +\frac{4}{3}\Theta_{\rm m} \sigma_{\rm m}^{ab} +2\,\sigma_{\rm m}^{c\langle a} \sigma_{{\rm m}c}^{b\rangle}
-\frac{2}{3}\, v^{\langle a} h_{\rm m}^{b\rangle c}\nabla_c \Theta_{\rm m}
+ \frac{4}{5}\,v^{\langle a} h_{\rm m}^{b\rangle c}\,h_{\rm m}^{de}\, \nabla_d\sigma_{ec}^{\rm m}\\
& \;\quad  - 3\,v_c\,\nabla^{\langle a} \sigma_{\rm m}^{bc \rangle} +O(v^2)\,, \\
\tilde{\xo}^{abc} &\circeq \nabla^{\langle a} \sigma_{\rm m}^{bc \rangle} + \frac{1}{3}\Big( 3\,\dot{\sigma}_{\rm m}^{\langle ab } -4\,\Theta_{\rm m} \sigma_{\rm m}^{\langle ab}-6\,\sigma_{\rm m}^{d\langle a} \sigma_{{\rm m}d}^{b}
\Big)v^{c\rangle} +O(v^2) \,.
\end{align}

\newpage
\subsection{Measurement of $\hoo^{-1}$ in bins of constant redshift}
\label{appc}

In this case, \eqref{ditoz} applies, together with its boosted version \eqref{tditoz}.
The disadvantage of this approach is that there is an infinite number of multipoles in the inverse Hubble parameter:
\begin{align} \label{hinv}
\frac{1}{\tho}\circeq 
\Big({\thsa +\tilde\hs_a\,\tn^a+ \tilde\hs_{ab}\,\tn^a \tn^b}\Big)^{-1} \circeq \sum_{\ell=0}^{\infty} \tilde{\ih}_{a_1a_2\cdots a_\ell}\, \tn^{\langle a_1} \tn^{a_2}\cdots \tn^{a_{\ell}\rangle}\,.
\end{align}
In order to find the relations obeyed by the infinite hierarchy of multipoles   in \eqref{hinv}, 
we need the following identities~\cite{1983AnPhy.150..455E, ETM1983,Gebbie:1998fe}:
\begin{align}
n^{\left(a_{\ell+1}\right.} \ih^{\left.A_{\ell}\right)} & = \ih^{A_{\ell+1}}+\frac{\ell^2}{(2 \ell+1)(2 \ell-1)}\,  h^{\left(a_{\ell+1} a_{\ell}\right.} \ih^{\left.A_{\ell-1}\right)}\,, \\
n^{\left(a_{\ell+2}\right.}n^{a_\ell+1} \ih^{\left.A_{\ell}\right)} &= \ih^{A_{\ell+2}}+\frac{2\ell^2+2\ell-1}{(2\ell+3)(2\ell-1)} h^{\left(a_{\ell+2} a_{\ell+1}\right.} 
\ih^{\left.A_{\ell}\right)} \notag\\
& +\frac{[\ell(\ell-1)]^2}{(2\ell+1)(2\ell-1)^2(2\ell-3)} h^{\left(a_{\ell+2} a_{\ell+1}\right.}h^{a_{\ell} a_{\ell-1}} \ih^{\left.A_{\ell-2}\right)}\,,
\end{align}
where we use the notation ${A_\ell}={a_1a_2\cdots a_\ell}$. 
These identities apply also to the boosted frame.
We multiply \eqref{hinv} by $\tilde\hs$
and then use the above identities.
For $\ell=0, 1$, we find
\begin{align}
\thsa\,\tiha +\frac{1}{3} \tilde\hs^a \,\tilde \ih_a+ \frac{4}{45}  \tilde \hs^{a b}\, \tilde \ih_{ab}
& =1, \\
\thsa \tilde \ih_a + \tilde \hs_a\, \tiha +\frac{2}{5}  \tilde \hs_{a b}\, \tilde\ih^b +\frac{6}{35} \tilde \hs^{b c}\,\tilde \ih_{abc} & =0,
\end{align}
and for $\ell \geq 2$,
\begin{align}
\thsa\, \tilde\ih_{A_{\ell}}+\tilde \ih_{\left\langle A_{\ell-2}\right.} \sigma^{\rm m}_{\left.a_{\ell-1} a_{\ell}\right\rangle}+\frac{2 \ell}{2 \ell+3}\, \tilde\ih_{b\left\langle A_{\ell-1}\right.} \sigma_{\left.a_{\ell}\right\rangle}^{{\rm m}\,b}+\frac{(\ell+1)(\ell+2)}{(2 \ell+3)(2 \ell+5)}\,\tilde \ih_{bc A_{\ell}} \sigma_{\rm m}^{bc}=0\,.
\end{align}

It is often sufficient to work to leading in relative velocity. In this case,
using \eqref{bhpobs}--\eqref{thhq} and  \eqref{nanbnc}, a Taylor expansion of \eqref{hinv} leads to
\begin{align} \label{hinvbf}
\tho^{-1}  \circeq {{\thsa}}^{-1} \bigg[ 1 & +
\Big(2\,v^a +\frac{2}{5}\, \hat\sigma_{\rm m}^{ab}\,v_b\Big) \tn_a
- \hat\sigma_{\rm m}^{ab} \tn_a\, \tn_b
- {4}\,v^{\langle a}\,\hat\sigma_{\rm m}^{bc\rangle}\, \tn_a \, \tn_b \, \tn_c
\bigg] +O(v^2)\,.
\end{align}
Here $\hat\sigma_{\rm m}^{ab}=\sigma_{\rm m}^{ab}/\hsa$ and $\thsa^{-1}=\hsa^{-1}+O(v^2)$ by \eqref{thhm}. In the matter frame, \eqref{hinvbf} reduces to
\begin{align}\label{hinvmf}
\ho^{-1} & \circeq \hsa^{-1}\Big[ 1- \hat\sigma_{\rm m}^{ab}\, n_a n_b
\Big] +O(v^2)\,.
\end{align}

It follows from \eqref{hinvbf} that at leading order in the boosted frame,  the boosted inverse Hubble parameter has a similar Doppler + aberration dipole and a similar quadrupole to the boosted Hubble parameter. However, it also has an octupole, $- 4\thsa^{-1} v^{\langle a}\sigma_{\rm m}^{bc\rangle}$, generated purely by relative motion. In the matter frame, \eqref{hinvmf} shows that the dipole and octupole vanish at leading order -- the same as for the Hubble parameter.

\newpage
\providecommand{\href}[2]{#2}\begingroup\raggedright
\endgroup


\begin{thebibliography}{10}
	
	\bibitem{Perivolaropoulos:2021jda}
	L.~Perivolaropoulos and F.~Skara, {\it {Challenges for \ensuremath{\Lambda}CDM: An update}},  {\em New Astron. Rev.} {\bf 95} (2022) 101659, [\href{http://arxiv.org/abs/2105.05208}{{\tt arXiv:2105.05208}}].
	
	\bibitem{Abdalla:2022yfr}
	E.~Abdalla et~al., {\it {Cosmology intertwined: A review of the particle physics, astrophysics, and cosmology associated with the cosmological tensions and anomalies}},  {\em JHEAp} {\bf 34} (2022) 49, [\href{http://arxiv.org/abs/2203.06142}{{\tt arXiv:2203.06142}}].
	
	\bibitem{Aluri:2022hzs}
	P.~K. Aluri et~al., {\it {Is the observable Universe consistent with the cosmological principle?}},  {\em Class. Quant. Grav.} {\bf 40} (2023) 094001, [\href{http://arxiv.org/abs/2207.05765}{{\tt arXiv:2207.05765}}].
	
	\bibitem{Peebles:2022akh}
	P.~J.~E. Peebles, {\it {Anomalies in physical cosmology}},  {\em Annals Phys.} {\bf 447} (2022) 169159, [\href{http://arxiv.org/abs/2208.05018}{{\tt arXiv:2208.05018}}].
	
	\bibitem{Verde:2023lmm}
	L.~Verde, N.~Sch\"oneberg, and H.~Gil-Mar\'\i{}n, {\it {A tale of many $H_0$}},  \href{http://arxiv.org/abs/2311.13305}{{\tt arXiv:2311.13305}}.
	
	\bibitem{Kalbouneh:2024tfw}
	B.~Kalbouneh, C.~Marinoni, and R.~Maartens, {\it {Cosmography of the local universe by multipole analysis of the expansion rate fluctuation field}},  \href{http://arxiv.org/abs/2401.12291}{{\tt arXiv:2401.12291}}.
	
	\bibitem{Ellis:1998ct}
	G.~F.~R. Ellis and H.~van Elst, {\it {Cosmological models: Cargese lectures 1998}},  {\em NATO Sci. Ser. C} {\bf 541} (1999) 1, [\href{http://arxiv.org/abs/gr-qc/9812046}{{\tt gr-qc/9812046}}].
	
	\bibitem{Tsagas:2007yx}
	C.~G. Tsagas, A.~Challinor, and R.~Maartens, {\it {Relativistic cosmology and large-scale structure}},  {\em Phys. Rept.} {\bf 465} (2008) 61, [\href{http://arxiv.org/abs/0705.4397}{{\tt arXiv:0705.4397}}].
	
	\bibitem{Clarkson:2010uz}
	C.~Clarkson and R.~Maartens, {\it {Inhomogeneity and the foundations of concordance cosmology}},  {\em Class. Quant. Grav.} {\bf 27} (2010) 124008, [\href{http://arxiv.org/abs/1005.2165}{{\tt arXiv:1005.2165}}].
	
	\bibitem{Ellis:2012}
	G.~F.~R. Ellis, R.~Maartens, and M.~A.~H. MacCallum, {\em {Relativistic Cosmology}}.
	\newblock Cambridge University Press, Cambridge, 2012.
	
	\bibitem{1983AnPhy.150..455E}
	G.~F.~R. {Ellis}, D.~R. {Matravers}, and R.~{Treciokas}, {\it {Anisotropic solutions of the Einstein-Boltzmann equations: I. General formalism}},  {\em Annals of Physics} {\bf 150} (1983) 455.
	
	\bibitem{ETM1983}
	G.~F.~R. {Ellis}, R.~{Treciokas}, and D.~R. {Matravers}, {\it {Anisotropic solutions of the Einstein-Boltzmann equations: II. Some exact properties of the equations}},  {\em Annals of Physics} {\bf 150} (1983) 487.
	
	\bibitem{Stoeger:1994qs}
	W.~R. Stoeger, R.~Maartens, and G.~F.~R. Ellis, {\it {Proving almost homogeneity of the universe: An Almost Ehlers-Geren-Sachs theorem}},  {\em Astrophys. J.} {\bf 443} (1995) 1.
	
	\bibitem{ellis1984}
	G.~Ellis and J.~Baldwin, {\it On the expected anisotropy of radio source counts},  {\em Mon. Not. Roy. Astron. Soc.} {\bf 206} (1984) 377.
	
	\bibitem{Maartens:1998xg}
	R.~Maartens, T.~Gebbie, and G.~F.~R. Ellis, {\it {Covariant cosmic microwave background anisotropies. 2. Nonlinear dynamics}},  {\em Phys. Rev. D} {\bf 59} (1999) 083506, [\href{http://arxiv.org/abs/astro-ph/9808163}{{\tt astro-ph/9808163}}].
	
	\bibitem{Kristian:1966}
	J.~Kristian and R.~K. Sachs, {\it {Observations in cosmology}},  {\em Astrophys. J.} {\bf 143} (1966) 379.
	
	\bibitem{Ellis:1971pg}
	G.~F.~R. Ellis, {\it {Relativistic cosmology}},  {\em Proc. Int. Sch. Phys. Fermi} {\bf 47} (1971) 104.
	
	\bibitem{ellis2009}
	G.~F.~R. Ellis, {\it Relativistic cosmology (reprint)},  {\em Gen. Relativ. Grav.} {\bf 41} (2009) 581.
	
	\bibitem{MacCallum:1970}
	M.~A.~H. MacCallum and G.~F.~R. Ellis, {\it {A Class of Homogeneous Cosmological Models. II~Observations}},  {\em Commun. math. Phys.} {\bf 19} (1970) 31.
	
	\bibitem{Maartens:1980}
	R.~Maartens, {\em {Idealised observations in relativistic cosmology}}.
	\newblock PhD thesis, University of Cape Town, 1980.
	
	\bibitem{Ellis:1985}
	G.~F.~R. Ellis, S.~D. Nel, R.~Maartens, W.~R. Stoeger, and A.~P. Whitman, {\it {Idealised observational cosmology}},  {\em Phys. Rep.} {\bf 124} (1985) 315.
	
	\bibitem{Clarkson:1999zq}
	C.~A. Clarkson, {\it {On the observational characteristics of inhomogeneous cosmologies}},  \href{http://arxiv.org/abs/astro-ph/0008089}{{\tt astro-ph/0008089}}.
	
	\bibitem{Umeh:2013}
	O.~Umeh, {\em {The influence of structure formation on the evolution of the Universe}}.
	\newblock PhD thesis, University of Cape Town, 2013.
	
	\bibitem{Heinesen:2020bej}
	A.~Heinesen, {\it {Multipole decomposition of the general luminosity distance 'Hubble law' -- a new framework for observational cosmology}},  {\em JCAP} {\bf 05} (2021) 008, [\href{http://arxiv.org/abs/2010.06534}{{\tt arXiv:2010.06534}}].
	
	\bibitem{Umeh:2022prn}
	O.~Umeh, {\it {Consequences of using a smooth cosmic distance in a lumpy universe. I.}},  {\em Phys. Rev. D} {\bf 106} (2022) 023514, [\href{http://arxiv.org/abs/2202.08230}{{\tt arXiv:2202.08230}}].
	
	\bibitem{Nadolny:2021hti}
	T.~Nadolny, R.~Durrer, M.~Kunz, and H.~Padmanabhan, {\it {A new way to test the Cosmological Principle: measuring our peculiar velocity and the large-scale anisotropy independently}},  {\em JCAP} {\bf 11} (2021) 009, [\href{http://arxiv.org/abs/2106.05284}{{\tt arXiv:2106.05284}}].
	
	\bibitem{Ivanov:2018cyw}
	D.~Ivanov, S.~Liberati, M.~Viel, and M.~Visser, {\it {Non-perturbative results for the luminosity and area distances}},  {\em JCAP} {\bf 06} (2018) 040, [\href{http://arxiv.org/abs/1802.06550}{{\tt arXiv:1802.06550}}].
	
	\bibitem{Maartens:2017qoa}
	R.~Maartens, C.~Clarkson, and S.~Chen, {\it {The kinematic dipole in galaxy redshift surveys}},  {\em JCAP} {\bf 01} (2018) 013, [\href{http://arxiv.org/abs/1709.04165}{{\tt arXiv:1709.04165}}].
	
	\bibitem{Sachs:1961zz}
	R.~K. Sachs, {\it {Gravitational waves in general relativity. 6. The outgoing radiation condition}},  {\em Proc. Roy. Soc. Lond. A} {\bf 264} (1961) 309.
	
	\bibitem{Hui:2005nm}
	L.~Hui and P.~B. Greene, {\it {Correlated Fluctuations in Luminosity Distance and the (Surprising) Importance of Peculiar Motion in Supernova Surveys}},  {\em Phys. Rev. D} {\bf 73} (2006) 123526, [\href{http://arxiv.org/abs/astro-ph/0512159}{{\tt astro-ph/0512159}}].
	
	\bibitem{Davis:2010jq}
	T.~M. Davis et~al., {\it {The Effect of Peculiar Velocities on Supernova Cosmology}},  {\em Astrophys. J.} {\bf 741} (2011) 67, [\href{http://arxiv.org/abs/1012.2912}{{\tt arXiv:1012.2912}}].
	
	\bibitem{Hasse:1999}
	W.~Hasse and V.~Perlick, {\it {On spacetime models with an isotropic Hubble flow}},  {\em Class. Quant. Grav.} {\bf 16} (1999) 2559.
	
	\bibitem{Visser:2003vq}
	M.~Visser, {\it {Jerk and the cosmological equation of state}},  {\em Class. Quant. Grav.} {\bf 21} (2004) 2603--2616, [\href{http://arxiv.org/abs/gr-qc/0309109}{{\tt gr-qc/0309109}}].
	
	\bibitem{Heinesen:2021azp}
	A.~Heinesen and H.~J. Macpherson, {\it {A prediction for anisotropies in the nearby Hubble flow}},  {\em JCAP} {\bf 03} (2022), no.~03 057, [\href{http://arxiv.org/abs/2111.14423}{{\tt arXiv:2111.14423}}].
	
	\bibitem{Kalbouneh:2022tfw}
	B.~Kalbouneh, C.~Marinoni, and J.~Bel, {\it {Multipole expansion of the local expansion rate}},  {\em Phys. Rev. D} {\bf 107} (2023) 023507, [\href{http://arxiv.org/abs/2210.11333}{{\tt arXiv:2210.11333}}].
	
	\bibitem{Umeh:2022hab}
	O.~Umeh, {\it {The art of building a smooth cosmic distance ladder in a perturbed universe}},  {\em JCAP} {\bf 08} (2022), no.~08 023, [\href{http://arxiv.org/abs/2201.11089}{{\tt arXiv:2201.11089}}].
	
	\bibitem{Umeh:2022kqs}
	O.~Umeh, {\it {Emergence of smooth distance and apparent magnitude in a lumpy Universe}},  {\em Class. Quant. Grav.} {\bf 39} (2022), no.~23 235006, [\href{http://arxiv.org/abs/2202.08237}{{\tt arXiv:2202.08237}}].
	
	\bibitem{Buchert:1997dr}
	T.~Buchert and A.~Dominguez, {\it {Modeling multistream flow in collisionless matter: approximations for large scale structure beyond shell crossing}},  {\em Astron. Astrophys.} {\bf 335} (1998) 395--402, [\href{http://arxiv.org/abs/astro-ph/9702139}{{\tt astro-ph/9702139}}].
	
	\bibitem{Lewis:2002nc}
	A.~Lewis and A.~Challinor, {\it {Evolution of cosmological dark matter perturbations}},  {\em Phys. Rev. D} {\bf 66} (2002) 023531, [\href{http://arxiv.org/abs/astro-ph/0203507}{{\tt astro-ph/0203507}}].
	
	\bibitem{Bartelmann:2019unp}
	M.~Bartelmann, I.~Kostyuk, E.~Kozlikin, R.~Lilow, C.~Littek, F.~Fabis, C.~Viermann, L.~Heisenberg, S.~Konrad, and D.~Geiss, {\it {Cosmic Structure Formation with Kinetic Field Theory}},  {\em Annalen Phys.} {\bf 531} (2019), no.~11 1800446, [\href{http://arxiv.org/abs/1905.01179}{{\tt arXiv:1905.01179}}].
	
	\bibitem{Fardeau:2023sxl}
	N.~Fardeau, T.~Buchert, F.~A. Roumi, and F.~Felegary, {\it {Nonperturbative collapse models for collisionless self-gravitating flows}},  {\em Phys. Rev. D} {\bf 108} (2023), no.~8 083502, [\href{http://arxiv.org/abs/2301.09154}{{\tt arXiv:2301.09154}}].
	
	\bibitem{Biagetti:2015hva}
	M.~Biagetti, V.~Desjacques, A.~Kehagias, D.~Racco, and A.~Riotto, {\it {The Halo Boltzmann Equation}},  {\em JCAP} {\bf 04} (2016) 040, [\href{http://arxiv.org/abs/1508.07330}{{\tt arXiv:1508.07330}}].
	
	\bibitem{Matsubara:2019tyb}
	T.~Matsubara, {\it {Velocity bias and the nonlinear perturbation theory of peaks}},  {\em Phys. Rev. D} {\bf 100} (2019), no.~8 083504, [\href{http://arxiv.org/abs/1907.13251}{{\tt arXiv:1907.13251}}].
	
	\bibitem{Clarkson:2011zq}
	C.~Clarkson, G.~Ellis, J.~Larena, and O.~Umeh, {\it {Does the growth of structure affect our dynamical models of the universe? The averaging, backreaction and fitting problems in cosmology}},  {\em Rept. Prog. Phys.} {\bf 74} (2011) 112901, [\href{http://arxiv.org/abs/1109.2314}{{\tt arXiv:1109.2314}}].
	
	\bibitem{Adamek:2017mzb}
	J.~Adamek, C.~Clarkson, D.~Daverio, R.~Durrer, and M.~Kunz, {\it {Safely smoothing spacetime: backreaction in relativistic cosmological simulations}},  {\em Class. Quant. Grav.} {\bf 36} (2019), no.~1 014001, [\href{http://arxiv.org/abs/1706.09309}{{\tt arXiv:1706.09309}}].
	
	\bibitem{Heinesen:2018vjp}
	A.~Heinesen, P.~Mourier, and T.~Buchert, {\it {On the covariance of scalar averaging and backreaction in relativistic inhomogeneous cosmology}},  {\em Class. Quant. Grav.} {\bf 36} (2019), no.~7 075001, [\href{http://arxiv.org/abs/1811.01374}{{\tt arXiv:1811.01374}}].
	
	\bibitem{Fanizza:2019pfp}
	G.~Fanizza, M.~Gasperini, G.~Marozzi, and G.~Veneziano, {\it {Generalized covariant prescriptions for averaging cosmological observables}},  {\em JCAP} {\bf 02} (2020) 017, [\href{http://arxiv.org/abs/1911.09469}{{\tt arXiv:1911.09469}}].
	
	\bibitem{Buchert:2022zaa}
	T.~Buchert, H.~van Elst, and A.~Heinesen, {\it {The averaging problem on the past null cone in inhomogeneous dust cosmologies}},  {\em Gen. Rel. Grav.} {\bf 55} (2023), no.~1 7, [\href{http://arxiv.org/abs/2202.10798}{{\tt arXiv:2202.10798}}].
	
	\bibitem{Wagner:2022etu}
	J.~Wagner, {\it {Solving the Hubble tension \`a la Ellis \& Stoeger 1987}},  {\em PoS} {\bf CORFU2022} (2023) 267, [\href{http://arxiv.org/abs/2203.11219}{{\tt arXiv:2203.11219}}].
	
	\bibitem{Anton:2023icm}
	T.~Anton and T.~Clifton, {\it {Modelling the emergence of cosmic anisotropy from non-linear structures}},  {\em Class. Quant. Grav.} {\bf 40} (2023), no.~14 145004, [\href{http://arxiv.org/abs/2302.05715}{{\tt arXiv:2302.05715}}].
	
	\bibitem{Gebbie:1998fe}
	T.~Gebbie and G.~F.~R. Ellis, {\it {Gauge-invariant covariant approach to cosmic background radiation anisotropies. Part 1.}},  {\em Annals Phys.} {\bf 282} (2006) 285, [\href{http://arxiv.org/abs/astro-ph/9804316}{{\tt astro-ph/9804316}}].
	
	\bibitem{Challinor:1998aa}
	A.~Challinor and A.~Lasenby, {\it {A Covariant and gauge invariant analysis of CMB anisotropies from scalar perturbations}},  {\em Phys. Rev. D} {\bf 58} (1998) 023001, [\href{http://arxiv.org/abs/astro-ph/9804150}{{\tt astro-ph/9804150}}].
	
\end{thebibliography}
\end{document}